\documentclass[12pt]{article}
\usepackage{amsmath}
\usepackage{graphicx,psfrag,epsf}
\usepackage{enumerate}

\usepackage{textcomp}
\usepackage[hyphens]{url} 
\usepackage{hyperref}

\newcommand{\blind}{0}

\newlength{\cslhangindent}
\newlength{\csllabelwidth}
\newlength{\cslentryspacingunit} 
  {}%
  {\par}
\newenvironment{CSLReferences}[2] 
 {
  \setlength{\parindent}{0pt}
  \ifodd #1
  \let\oldpar\par
  \def\par{\hangindent=\cslhangindent\oldpar}
  \fi
  \setlength{\parskip}{#2\cslentryspacingunit}
 }%
 {}
\usepackage{calc}

\usepackage{amsthm}
\newtheorem{thm}{Theorem}

\usepackage{booktabs}
\usepackage{longtable}
\usepackage[natbibapa]{apacite}
\usepackage{caption}
\usepackage{multirow}
\usepackage{float}    
\usepackage{rotfloat} 
\usepackage{pdflscape}
\hypersetup{hidelinks}

\begin{document}

\def\spacingset#1{\renewcommand{\baselinestretch}%
{#1}\small\normalsize} \spacingset{1}


\if0\blind
{
  \title{\bf Small sample methods for cluster-robust variance estimation
and hypothesis testing in fixed effects models}

  \author{
        James E. Pustejovsky \thanks{Department of Educational
Psychology, University of Wisconsin - Madison, 1025 West Johnson Street,
Madison, WI 53706. Email:
\href{mailto:pustejovsky@wisc.edu}{\nolinkurl{pustejovsky@wisc.edu}}} \\
    University of Wisconsin - Madison\\
     and \\     Elizabeth Tipton \thanks{Department of Statistics,
Northwestern University. Email:
\href{mailto:tipton@northwestern.edu}{\nolinkurl{tipton@northwestern.edu}}} \\
    Northwestern University\\
      }
  \maketitle
} \fi

\if1\blind
{
  \bigskip
  \bigskip
  \bigskip
  \begin{center}
    {\LARGE\bf Small sample methods for cluster-robust variance
estimation and hypothesis testing in fixed effects models}
  \end{center}
  \medskip
} \fi

\bigskip
\begin{abstract}
In panel data models and other regressions with unobserved effects,
fixed effects estimation is often paired with cluster-robust variance
estimation (CRVE) in order to account for heteroskedasticity and
un-modeled dependence among the errors. Although asymptotically
consistent, CRVE can be biased downward when the number of clusters is
small, leading to hypothesis tests with rejection rates that are too
high. More accurate tests can be constructed using bias-reduced
linearization (BRL), which corrects the CRVE based on a working model,
in conjunction with a Satterthwaite approximation for t-tests. We
propose a generalization of BRL that can be applied in models with
arbitrary sets of fixed effects, where the original BRL method is
undefined, and describe how to apply the method when the regression is
estimated after absorbing the fixed effects. We also propose a
small-sample test for multiple-parameter hypotheses, which generalizes
the Satterthwaite approximation for t-tests. In simulations covering a
wide range of scenarios, we find that the conventional cluster-robust
Wald test can severely over-reject while the proposed small-sample test
maintains Type I error close to nominal levels. The proposed methods are
implemented in an R package called clubSandwich.

\textbf{This paper has been revised to correct an error in the statement
of Theorem 2 and minor typos in the proof of Theorem 1.}
\end{abstract}

\noindent%
\vfill

\newpage
\spacingset{1.45} 

\hypertarget{sec:intro}{%
\section{INTRODUCTION}\label{sec:intro}}

In many economic analyses, interest centers on the parameters of a
linear regression model estimated by ordinary or weighted least squares
from data exhibiting within-group dependence. Such dependence can arise
from sampling or random assignment of aggregate units (e.g., counties,
districts, villages), each of which contains multiple observations; from
repeated measurement of an outcome on a common set of units, as in panel
data; or from model misspecification, as in analysis of regression
discontinuity designs (e.g., Lee and Card 2008). A common approach to
inference in these settings is to use a cluster-robust variance
estimator (CRVE, Arellano 1987; Liang and Zeger 1986; White 1984). The
advantage of the CRVE is that it produces consistent standard errors and
test statistics without imposing strong parametric assumptions about the
correlation structure of the errors in the model. Instead, the method
relies on the weaker assumption that units can be grouped into clusters
that are mutually independent. In the past decade, use of CRVEs has
become standard practice for micro-economic researchers, as evidenced by
coverage in major textbooks and review articles (e.g., Wooldridge 2010;
Angrist and Pischke 2009; Cameron and Miller 2015).

As a leading example, consider a difference-in-differences analysis of
state-by-year panel data, where the goal is to understand the effects on
employment outcomes of several state-level policy shifts. Each policy
effect would be parameterized as a dummy variable in a regression model,
which might also include other demographic controls. It is also common
to include fixed effects for states and time-points in order to control
for unobserved confounding in each dimension. The model could be
estimated by least squares with the fixed effects included as dummy
variables (or what we will call the LSDV estimator). More commonly, the
effects of the policy indicators would be estimated after absorbing the
fixed effects, a computational technique that is also known as the fixed
effects estimator or ``within transformation'' (Wooldridge 2010).
Standard errors would then be clustered by state to account for residual
dependence in the errors from a given state, and these clustered
standard errors would be used to test hypotheses about the set of
policies. The need to cluster the standard errors by state, even when
including state fixed effects, was highlighted by Bertrand, Duflo, and
Mullainathan (2004), who showed that to do otherwise can lead to
inappropriately small standard errors and hypothesis tests with
incorrect rejection rates.

The consistency property of CRVEs is asymptotic in the number of
independent clusters (Wooldridge 2003). Recent methodological work has
demonstrated that CRVEs can be biased downward and associated hypothesis
tests can have Type-I error rates considerably in excess of nominal
levels when based on a small or moderate number of clusters (e.g.,
MacKinnon and Webb 2016). Cameron and Miller (2015) provided an
extensive review of this literature, including a discussion of current
practice, possible solutions, and open problems. In particular, they
demonstrated that small-sample corrections for t-tests implemented in
common software packages such as Stata and SAS do not provide adequate
control of Type-I error.

Bell and McCaffrey (2002) proposed a method that improves the
small-sample properties of CRVEs (see also McCaffrey, Bell, and Botts
2001). The method, called bias-reduced linearization (BRL), entails
adjusting the CRVE so that it is exactly unbiased under a working model
specified by the analyst, while also remaining asymptotically consistent
under arbitrary true variance structures. Simulations reported by Bell
and McCaffrey (2002) demonstrate that the BRL correction serves to
reduce the bias of the CRVE even when the working model is misspecified.
The same authors also proposed small-sample corrections to
single-parameter hypothesis tests using the BRL variance estimator,
based on Satterthwaite (Bell and McCaffrey 2002) or saddlepoint
approximations (McCaffrey and Bell 2006). In an analysis of a
longitudinal cluster-randomized trial with 35 clusters, Angrist and Lavy
(2009) observed that the BRL correction makes a difference for
inferences.

Despite a growing body of evidence that BRL performs well (e.g., Imbens
and Kolesar 2016), several problems with the method hinder its wider
application. First, Angrist and Pischke (2009) noted that the BRL
correction is undefined in some commonly used models, such as
state-by-year panels that include fixed effects for states and for years
(see also Young 2016). Second, in models with fixed effects, the
magnitude of the BRL adjustment depends on whether it is computed based
on the full design matrix (i.e., the LSDV estimator) or after absorbing
the fixed effects. Third, extant methods for hypothesis testing based on
BRL are limited to single-parameter constraints (Bell and McCaffrey
2002; McCaffrey and Bell 2006) and small-sample methods for
multiple-parameter hypothesis tests remain lacking.

This paper addresses each of these concerns in turn, with the aim of
extending the BRL method so that is suitable for general application.
First, we describe a simple modification of the BRL adjustment that
remains well-defined in models with arbitrary sets of fixed effects,
where existing BRL adjustments break down. Second, we demonstrate how to
calculate the BRL adjustments based on the fixed effects estimator and
identify conditions under which first-stage absorption of the fixed
effects can be ignored. Finally, we propose a procedure for testing
multiple-parameter hypotheses by approximating the sampling distribution
of the Wald statistic using Hotelling's \(T^2\) distribution with
estimated degrees of freedom. The method is a generalization of the
Satterthwaite correction proposed by Bell and McCaffrey (2002) for
single parameter constraints. The proposed methods are implemented in
the R package \texttt{clubSandwich}, which is available on the
Comprehensive R Archive Network.

Our work is related to a stream of recent literature that has examined
methods for cluster-robust inference with a small number of clusters.
Conley and Taber (2011) proposed methods for hypothesis testing in a
difference-in-differences setting where the number of treated units is
small and fixed, while the number of untreated units increases
asymptotically. Ibragimov and Müller (2010, 2016) proposed
cluster-robust t-tests that maintain the nominal Type-I error rate by
re-weighting within-cluster estimators of the target parameter. Young
(2016) proposed a Satterthwaite correction for t-tests based on a
different type of bias correction to the CRVE, where the bias correction
term is derived under a working model. Cameron, Gelbach, and Miller
(2008) investigated a range of bootstrapping procedures that provide
improved Type-I error control in small samples, finding that a cluster
wild-bootstrap technique was particularly accurate in small samples.
Nearly all of this work has focused on single-parameter hypothesis tests
only. For multiple-parameter constraints, Cameron and Miller (2015)
suggested an ad hoc degrees of freedom adjustment and noted, as an
alternative, that bootstrapping techniques can in principle be applied
to multiple-parameter tests. However, little methodological work has
examined the accuracy of multiple-parameter tests.

The paper is organized as follows. The remainder of this section
introduces our econometric framework and reviews standard CRVE methods,
as implemented in most software applications. Section \ref{sec:BRL}
reviews the original BRL correction and describes modifications that
make it possible to implement BRL in a broad class of models with fixed
effects. Section \ref{sec:testing} discusses hypothesis tests based on
the BRL-adjusted CRVE. Section \ref{sec:simulation} reports a simulation
study examining the null rejection rates of multiple-parameter
hypothesis tests, where we find that the small-sample test offers
drastic improvements over commonly implemented alternatives. Section
\ref{sec:examples} illustrates the use of the proposed hypothesis tests
in two applications. Section \ref{sec:conclusion} concludes and
discusses avenues for future work.

\hypertarget{econometric-framework}{%
\subsection{Econometric framework}\label{econometric-framework}}

We consider a linear regression model of the form, \begin{equation}
\label{eq:fixed_effects}
\mathbf{y}_i = \mathbf{R}_i \boldsymbol\beta + \mathbf{S}_i \boldsymbol\gamma + \mathbf{T}_i \boldsymbol\mu + \boldsymbol\epsilon_i,
\end{equation} where there are a total of \(m\) clusters; cluster \(i\)
contains \(n_i\) units; \(\mathbf{y}_i\) is a vector of the \(n_i\)
values of the outcome for units in cluster \(i\); \(\mathbf{R}_i\) is an
\(n_i \times r\) matrix containing predictors of primary interest (e.g.,
policy variables) and any additional controls; \(\mathbf{S}_i\) is an
\(n_i \times s\) matrix describing fixed effects that are identified
across multiple clusters; and \(\mathbf{T}_i\) is an \(n_i \times t\)
matrix describing cluster-specific fixed effects, which must satisfy
\(\mathbf{T}_h \mathbf{T}_i' = \mathbf{0}\) for \(h \neq i\). Note that
the distinction between the covariates \(\mathbf{R}_i\) versus the fixed
effects \(\mathbf{S}_i\) is arbitrary and depends on the analyst's
inferential goals. In a fixed effects model for state-by-year panel
data, \(\mathbf{R}_i\) would include variables describing policy
changes, as well as additional demographic controls; \(\mathbf{S}_i\)
would include year fixed effects; and \(\mathbf{T}_i\) would indicate
state fixed effects (and perhaps also state-specific time trends).
Interest would center on testing hypotheses regarding the coefficients
in \(\boldsymbol\beta\) that correspond to the policy indicators, while
\(\boldsymbol\gamma\) and \(\boldsymbol\mu\) would be treated as
incidental.

We shall assume that
\(\text{E}\left(\boldsymbol\epsilon_i\left|\mathbf{R}_i,\mathbf{S}_i, \mathbf{T}_i\right.\right) = \mathbf{0}\)
and
\(\text{Var}\left(\boldsymbol\epsilon_i\left|\mathbf{R}_i,\mathbf{S}_i,\mathbf{T}_i\right.\right) = \boldsymbol\Sigma_i\),
for \(i = 1,...,m\), where the form of
\(\boldsymbol\Sigma_1,...,\boldsymbol\Sigma_m\) may be unknown but the
errors are independent across clusters. Let
\(\mathbf{U}_i = \left[\mathbf{R}_i \ \mathbf{S}_i \right]\) denote the
set of predictors that are identified across multiple clusters,
\(\mathbf{X}_i = \left[\mathbf{R}_i \ \mathbf{S}_i \ \mathbf{T}_i \right]\)
denote the full set of predictors,
\(\boldsymbol\alpha = \left(\boldsymbol\beta', \boldsymbol\gamma', \boldsymbol\mu' \right)'\),
and \(p = r + s + t\). Let \(N = \sum_{i=1}^m n_i\) denote the total
number of observations. Let \(\mathbf{y}\), \(\mathbf{R}\),
\(\mathbf{S}\), \(\mathbf{T}\), \(\mathbf{U}\), \(\mathbf{X}\), and
\(\boldsymbol\epsilon\) denote the matrices obtained by stacking their
corresponding components, as in
\(\mathbf{R} = \left(\mathbf{R}_1' \ \mathbf{R}_2' \ \cdots \ \mathbf{R}_m'\right)'\).

We assume that \(\boldsymbol\beta\) is estimated by weighted least
squares (WLS) using symmetric, full rank weighting matrices
\(\mathbf{W}_1,...,\mathbf{W}_m\). Clearly, the WLS estimator includes
ordinary least squares (OLS) as a special case. More generally, the WLS
estimator encompasses feasible generalized least squares, where it is
assumed that
\(\text{Var}\left(\mathbf{e}_i\left|\mathbf{X}_i\right.\right) = \boldsymbol\Phi_i\),
a known function of a low-dimensional parameter. For example, an
auto-regressive error structure might be posited to describe repeated
measures on an individual over time. The weighting matrices are then
taken to be \(\mathbf{W}_i = \hat{\boldsymbol\Phi}_i^{-1}\), where the
\(\hat{\boldsymbol\Phi}_i\) are constructed from estimates of the
variance parameter. Finally, for analysis of data from complex survey
designs, WLS may be used with sampling weights in order to account for
unequal selection probabilities.

\hypertarget{absorption}{%
\subsection{Absorption}\label{absorption}}

The goal of most analyses is to estimate and test hypotheses regarding
the parameters in \(\boldsymbol\beta\), while the fixed effects
\(\boldsymbol\gamma\) and \(\boldsymbol\mu\) are not of inferential
interest. Furthermore, LSDV estimation becomes computationally intensive
and numerically inaccurate if the model includes a large number of fixed
effects (i.e., \(s + t\) large). A commonly implemented alternative to
LSDV is to first absorb the fixed effects, which leaves only the \(r\)
parameters in \(\boldsymbol\beta\) to be estimated. Because Section
\ref{sec:BRL} examines the implications of absorption for application of
the BRL adjustment, we now formalize this procedure. Denote the full
block-diagonal weighting matrix as
\(\mathbf{W} = \text{diag}\left(\mathbf{W}_1,...,\mathbf{W}_m\right)\).
Let \(\mathbf{K}\) be the \(p \times r\) matrix that selects the
covariates of interest, so that \(\mathbf{X} \mathbf{K} = \mathbf{R}\)
and \(\mathbf{K}'\boldsymbol\alpha = \boldsymbol\beta\). For a generic
matrix \(\mathbf{Z}\) of full column rank, let
\(\mathbf{M_Z} = \left(\mathbf{Z}'\mathbf{W}\mathbf{Z}\right)^{-1}\) and
\(\mathbf{H_Z} = \mathbf{Z}\mathbf{M_Z}\mathbf{Z}'\mathbf{W}\).

The absorption technique involves obtaining the residuals from the
regression of \(\mathbf{y}\) on \(\mathbf{T}\) and from the multivariate
regression of \([\mathbf{R} \ \mathbf{S}]\) on \(\mathbf{T}\). The
\(\mathbf{y}\) residuals and \(\mathbf{R}\) residuals are then regressed
on the \(\mathbf{S}\) residuals. Finally, these twice-regressed
\(\mathbf{y}\) residuals are regressed on the twice-regressed
\(\mathbf{R}\) residuals to obtain the WLS estimates of
\(\boldsymbol\beta\). Let
\(\mathbf{\ddot{S}} = \left(\mathbf{I} - \mathbf{H_T}\right)\mathbf{S}\),
\(\mathbf{\ddot{R}} = \left(\mathbf{I} - \mathbf{H_{\ddot{S}}}\right)\left(\mathbf{I} - \mathbf{H_T}\right)\mathbf{R}\),
and
\(\mathbf{\ddot{y}} = \left(\mathbf{I} - \mathbf{H_{\ddot{S}}}\right)\left(\mathbf{I} - \mathbf{H_T}\right)\mathbf{y}\).
In what follows, subscripts on \(\mathbf{\ddot{R}}\),
\(\mathbf{\ddot{S}}\), \(\mathbf{\ddot{U}}\), and \(\mathbf{\ddot{y}}\)
refer to the rows of these matrices corresponding to a specific cluster.
The WLS estimator of \(\boldsymbol\beta\) can then be written as
\begin{equation}
\label{eq:WLS}
\boldsymbol{\hat\beta} = \mathbf{M_{\ddot{R}}} \sum_{i=1}^m \mathbf{\ddot{R}}_i' \mathbf{W}_i \mathbf{\ddot{y}}_i. 
\end{equation} This estimator is algebraically identical to the LSDV
estimator,
\(\boldsymbol{\hat\beta} = \mathbf{K}'\mathbf{M_X} \mathbf{X}' \mathbf{W} \mathbf{y}\),
but avoids the need to solve a system of \(p\) linear equations. For
further details on sequential absorption, see Davis (2002). In the
remainder, we assume that fixed effects are absorbed before estimation
of \(\boldsymbol\beta\).

\hypertarget{standard-crve}{%
\subsection{Standard CRVE}\label{standard-crve}}

The WLS estimator \(\boldsymbol{\hat\beta}\), has true variance
\begin{equation}
\label{eq:var_WLS}
\text{Var}\left(\boldsymbol{\hat\beta}\right) = \mathbf{M_{\ddot{R}}}\left(\sum_{i=1}^m \mathbf{\ddot{R}}_i' \mathbf{W}_i \boldsymbol\Sigma_i \mathbf{W}_i\mathbf{\ddot{R}}_i\right) \mathbf{M_{\ddot{R}}},
\end{equation} which depends upon the unknown variance matrices
\(\boldsymbol\Sigma_i\).

The CRVE involves estimating
\(\text{Var}\left(\boldsymbol{\hat\beta}\right)\) empirically, without
imposing structural assumptions on \(\boldsymbol\Sigma_i\). There are
several versions of this approach, all of which can be written as
\begin{equation}
\label{eq:V_small}
\mathbf{V}^{CR} = \mathbf{M_{\ddot{R}}}\left(\sum_{i=1}^m \mathbf{\ddot{R}}_i'\mathbf{W}_i \mathbf{A}_i \mathbf{e}_i \mathbf{e}_i' \mathbf{A}_i' \mathbf{W}_i \mathbf{\ddot{R}}_i\right) \mathbf{M_{\ddot{R}}},
\end{equation} where
\(\mathbf{e}_i = \mathbf{Y}_i - \mathbf{X}_i \boldsymbol{\hat\beta}\) is
the vector of residuals from cluster \(i\) and \(\mathbf{A}_i\) is some
\(n_i\) by \(n_i\) adjustment matrix.

The form of the adjustment matrices parallels those of the
heteroskedasticity-consistent variance estimators proposed by MacKinnon
and White (1985). The original CRVE, described by Liang and Zeger
(1986), uses \(\mathbf{A}_i = \mathbf{I}_i\), an \(n_i \times n_i\)
identity matrix. Following Cameron and Miller (2015), we refer to this
estimator as CR0. This estimator is biased towards zero because the
cross-product of the residuals \(\mathbf{e}_i \mathbf{e}_i'\) tends to
under-estimate the true variance \(\boldsymbol\Sigma_i\) in cluster
\(i\). A rough bias adjustment is to take
\(\mathbf{A}_i = c\mathbf{I}_i\), where \(c = \sqrt{(m/(m-1))}\); we
denote this adjusted estimator as CR1. Some functions in Stata use a
slightly different correction factor
\(c_S = \sqrt{(m N)/[(m - 1)(N - p)]}\); we will refer to the adjusted
estimator using \(c_S\) as CR1S. When \(N >> p\),
\(c_S \approx \sqrt{m/(m-1)}\) and so CR1 and CR1S will be very similar.
However, CR1 and CR1S can differ quite substantially for models with a
large number of fixed effects and small within-cluster sample size;
recent guidance emphasizes that CR1S is not appropriate for this
scenario (Cameron and Miller 2015). The CR1 and CR1S estimators are
commonly used in empirical applications.

Use of these adjustments still tends to under-estimate the true variance
of \(\hat{\boldsymbol\beta}\) because the degree of bias depends not
only on the number of clusters \(m\), but also on skewness of the
covariates and unbalance across clusters (Carter, Schnepel, and
Steigerwald 2013; MacKinnon 2013; Cameron and Miller 2015; Young 2016).
A more principled approach to bias correction would take into account
the features of the covariates in \(\mathbf{X}\). One such estimator
uses adjustment matrices given by
\(\mathbf{A}_i = \left(\mathbf{I} - \mathbf{\ddot{R}}_i \mathbf{M_{\ddot{R}}}\mathbf{\ddot{R}}_i'\mathbf{W}_i\right)^{-1}\).
This estimator, denoted CR3, closely approximates the jackknife
re-sampling estimator (Bell and McCaffrey 2002; Mancl and DeRouen 2001).
However, CR3 tends to over-correct the bias of CR0, while the CR1
estimator tends to under-correct. The next section describes in detail
the BRL approach, which makes adjustments that are intermediate in
magnitude between CR1 and CR3.

\hypertarget{sec:BRL}{%
\section{BIAS REDUCED LINEARIZATION}\label{sec:BRL}}

The BRL correction is premised on a ``working'' model for the structure
of the errors, which must be specified by the analyst. Under a given
working model, adjustment matrices \(\mathbf{A}_i\) are defined so that
the variance estimator is exactly unbiased. We refer to this correction
as CR2 because it extends the HC2 variance estimator for regressions
with uncorrelated errors, which is exactly unbiased when the errors are
homoskedastic (MacKinnon and White 1985). The idea of specifying a model
may seem antithetical to the purpose of using CRVE, yet extensive
simulation studies have demonstrated that the method performs well in
small samples even when the working model is incorrect (Tipton 2015;
Bell and McCaffrey 2002; Cameron and Miller 2015; Imbens and Kolesar
2016). Although the CR2 estimator might not be exactly unbiased when the
working model is misspecified, its bias still tends to be greatly
reduced compared to CR1 or CR0 (thus the name ``bias reduced
linearization''). Furthermore, as the number of clusters increases,
reliance on the working model diminishes.

Let
\(\boldsymbol\Phi = \text{diag}\left(\boldsymbol\Phi_1,...,\boldsymbol\Phi_m\right)\)
denote a working model for the covariance structure (up to a scalar
constant). For example, we might assume that the errors are uncorrelated
and homoskedastic, with \(\boldsymbol\Phi_i = \mathbf{I}_i\) for
\(i = 1,...,m\). Alternatively, Imbens and Kolesar (2016) suggested
using a random effects (i.e., compound symmetric) structure, in which
\(\boldsymbol\Phi_i\) has unit diagonal entries and off-diagonal entries
of \(\rho\), with \(\rho\) estimated using the OLS residuals.

In the original formulation of Bell and McCaffrey (2002), the BRL
adjustment matrices are chosen to satisfy the criterion \begin{equation}
\label{eq:CR2_criterion_BM}
\mathbf{A}_i \left(\mathbf{I} - \mathbf{H_X}\right)_i \boldsymbol\Phi \left(\mathbf{I} - \mathbf{H_X}\right)_i' \mathbf{A}_i'  =  \boldsymbol\Phi_i 
\end{equation} for a given working model, where
\(\left(\mathbf{I} - \mathbf{H_X}\right)_i\) denotes the rows of
\(\mathbf{I} - \mathbf{H_X}\) corresponding to cluster \(i\). If the
working model and weight matrices are both taken to be identity
matrices, then the adjustment matrices simplify to
\(\mathbf{A}_i = \left(\mathbf{I}_i - \mathbf{X}_i \mathbf{M_{X}} \mathbf{X}_i'\right)^{-1/2}\),
where \(\mathbf{Z}^{-1/2}\) denotes the symmetric square-root of the
matrix \(\mathbf{Z}\).

\hypertarget{a-more-general-brl-criterion}{%
\subsection{A more general BRL
criterion}\label{a-more-general-brl-criterion}}

The original formulation of \(\mathbf{A}_i\) is problematic because, for
some fixed effects models that are common in economic applications,
Equation \ref{eq:CR2_criterion_BM} has no solution. Angrist and Pischke
(2009) note that this problem occurs in balanced state-by-year panel
models that include fixed effects for states and for years, where
\(\mathbf{I}_i - \mathbf{X}_i \mathbf{M_{X}} \mathbf{X}_i'\) is not of
full rank. Young (2016) reported that this problem occurred frequently
when applying BRL to a large corpus of fitted regression models drawn
from published studies.

This issue can be solved by using an alternative criterion to define the
adjustment matrices, for which a solution always exists. Instead of
(\ref{eq:CR2_criterion_BM}), we propose to use adjustment matrices
\(\mathbf{A}_i\) that satisfy: \begin{equation}
\label{eq:CR2_criterion}
\mathbf{\ddot{R}}_i' \mathbf{W}_i \mathbf{A}_i \left(\mathbf{I} - \mathbf{H_X}\right)_i \boldsymbol\Phi \left(\mathbf{I} - \mathbf{H_X}\right)_i' \mathbf{A}_i' \mathbf{W}_i \mathbf{\ddot{R}}_i = \mathbf{\ddot{R}}_i' \mathbf{W}_i \boldsymbol\Phi_i \mathbf{W}_i \mathbf{\ddot{R}}_i.
\end{equation} A variance estimator that uses such adjustment matrices
will be exactly unbiased when the working model is correctly specified.

A symmetric solution to Equation (\ref{eq:CR2_criterion}) is given by
\begin{equation}
\label{eq:CR2_adjustment}
\mathbf{A}_i = \mathbf{D}_i' \mathbf{B}_i^{+1/2} \mathbf{D}_i,
\end{equation} where \(\mathbf{D}_i\) is the upper-right triangular
Cholesky factorization of \(\boldsymbol\Phi_i\), \begin{equation}
\label{eq:CR2_Bmatrix}
\mathbf{B}_i = \mathbf{D}_i\left(\mathbf{I} - \mathbf{H_X}\right)_i \boldsymbol\Phi \left(\mathbf{I} - \mathbf{H_X}\right)_i' \mathbf{D}_i',
\end{equation} and \(\mathbf{B}_i^{+1/2}\) is the symmetric square root
of the Moore-Penrose inverse of \(\mathbf{B}_i\). The Moore-Penrose
inverse of \(\mathbf{B}_i\) is well-defined and unique (Banerjee and Roy
2014, Thm. 9.18). In contrast, the original BRL adjustment matrices
involve the symmetric square root of the regular inverse of
\(\mathbf{B}_i\), which does not exist when \(\mathbf{B}_i\) is
rank-deficient. If \(\mathbf{B}_i\) is of full rank, then our adjustment
matrices reduce to the original formulation described by Bell and
McCaffrey (2002).

The adjustment matrices given by (\ref{eq:CR2_adjustment}) and
(\ref{eq:CR2_Bmatrix}) satisfy criterion (\ref{eq:CR2_criterion}), as
stated in the following theorem.

\begin{thm}
\label{thm:BRL_FE}
Let $\mathbf{L}_i = \left(\mathbf{\ddot{U}}'\mathbf{W}\mathbf{\ddot{U}} - \mathbf{\ddot{U}}_i'\mathbf{W}_i\mathbf{\ddot{U}}_i\right)$, where $\mathbf{\ddot{U}} = \left(\mathbf{I} - \mathbf{H_T}\right)\mathbf{U}$, and assume that $\mathbf{L}_1,...,\mathbf{L}_m$ have full rank $r + s$. Further assume that $\text{Var}\left(\boldsymbol\epsilon_i\left|\mathbf{X}_i\right.\right) = \sigma^2 \boldsymbol\Phi_i$, for $i = 1,...,m$. Then the adjustment matrix $\mathbf{A}_i$ defined in (\ref{eq:CR2_adjustment}) and (\ref{eq:CR2_Bmatrix}) satisfies criterion (\ref{eq:CR2_criterion}) and the CR2 variance estimator is exactly unbiased.
\end{thm}

Proof is given in \ref{app:proof1}. The main implication of Theorem
\ref{thm:BRL_FE} is that, under our more general definition, the CR2
variance estimator remains well-defined even in models with large sets
of fixed effects.

\hypertarget{absorption-and-lsdv-equivalence}{%
\subsection{Absorption and LSDV
Equivalence}\label{absorption-and-lsdv-equivalence}}

In fixed effects regression models, a problem with the original
definition of BRL is that it can result in a different estimator
depending upon which design matrix is used. If \(\boldsymbol\beta\) is
estimated using LSDV, then it is natural to calculate the CR2 adjustment
matrices based on the full covariate design matrix, \(\mathbf{X}\).
However, if \(\boldsymbol\beta\) is estimated after absorbing the fixed
effects, the analyst might choose to calculate the CR2 correction based
on the absorbed covariate matrix \(\mathbf{\ddot{R}}\)---that is, by
substituting \(\mathbf{H_{\ddot{R}}}\) for \(\mathbf{H_X}\) in
(\ref{eq:CR2_Bmatrix})---in order to avoid calculating the full
projection matrix \(\mathbf{H_X}\). This approach can lead to different
adjustment matrices because it is based on a subtly different working
model. Essentially, calculating CR2 based on \(\mathbf{H_{\ddot{R}}}\)
amounts to assuming that the working model \(\boldsymbol\Phi\) applies
not to the model errors \(\boldsymbol\epsilon\), but rather to the
errors from the final-stage regression of \(\mathbf{\ddot{y}}\) on
\(\mathbf{\ddot{R}}\). Because the CR2 adjustment is relatively
insensitive to the working model, the difference between accounting for
or ignoring absorption will in many instances be small. We investigate
the differences between the approaches as part of the simulation study
in Section \ref{sec:simulation}.

When based on the full regression model, a drawback of using the CR2
adjustment matrices is that it entails calculating the projection matrix
\(\mathbf{H_X}\) for the full set of \(p\) covariates (i.e., including
fixed effect indicators). Given that the entire advantage of using
absorption to calculate \(\hat{\boldsymbol\beta}\) is to avoid
computations involving large, sparse matrices, it is of interest to find
methods for more efficiently calculating the CR2 adjustment matrices.
Some computational efficiency can be gained by using the fact that the
residual projection matrix \(\mathbf{I} - \mathbf{H_X}\) can be factored
into components as
\(\left(\mathbf{I} - \mathbf{H_X}\right)_i = \left(\mathbf{I} - \mathbf{H_{\ddot{R}}}\right)_i \left(\mathbf{I} - \mathbf{H_{\ddot{S}}}\right) \left(\mathbf{I} - \mathbf{H_T}\right)\).

In certain circumstances, further computational efficiency can be
achieved by computing the adjustment matrices after absorbing the
within-cluster fixed effects \(\mathbf{T}\) (but not the between-cluster
fixed effects \(\mathbf{S}\)). Specifically, for ordinary (unweighted)
least squares estimation with an identity working model
\((\boldsymbol\Phi = \mathbf{I})\), the adjustment matrices can be
calculated without accounting for the within-cluster fixed effects, as
\begin{equation}
\label{eq:CR2_tilde}
\mathbf{\tilde{A}}_i = \left(\mathbf{I}_i - \mathbf{\ddot{U}}_i\mathbf{M}_{\mathbf{\ddot{U}}} \mathbf{\ddot{U}}_i'\right)^{-1/2}
\end{equation} This result is formalized in the following
theorem.\footnote{In a previous version of the paper, Theorem
  \ref{thm:absorb} made a more general claim about the equivalence of
  the adjustment matrices calculated with or without accounting for the
  within-cluster fixed effects. The previous statement of the theorem
  asserted that the equivalence holds for weighted least squares so long
  as the weights are inverse of the working model,
  \(\mathbf{W}_i = \boldsymbol\Phi^{-1}\) for \(i = 1,..,m\). However,
  the claimed equivalence does not hold generally. The proof of the
  previous statement of Theorem \ref{thm:absorb} relied on a Woodbury
  identity for generalized inverses that does not hold for
  \(\mathbf{B}_i\) because necessary rank conditions are not satisfied.}

\begin{thm}
\label{thm:absorb}
Let $\mathbf{\tilde{V}}^{CR}$ be the CRVE calculated using $\mathbf{\tilde{A}}_i$ as defined in (\ref{eq:CR2_tilde}) and $\mathbf{V}^{CR}$ be calculated using the full adjustment matrices as defined in (\ref{eq:CR2_adjustment}) and (\ref{eq:CR2_Bmatrix}).
If $\mathbf{W}_i = \mathbf{I}_i$ and $\boldsymbol\Phi_i = \mathbf{I}_i$ for $i = 1,...,m$, then $\mathbf{A}_i \mathbf{\ddot{R}}_i = \mathbf{\tilde{A}}_i \mathbf{\ddot{R}}_i$ and $\mathbf{\tilde{V}}^{CR} = \mathbf{V}^{CR}$.
\end{thm}

Proof is given in \ref{app:proof2}. The main implication of Theorem
\ref{thm:absorb} is that the more computationally convenient formula
\(\mathbf{\tilde{A}}_i\) can be used in the common case of unweighted
least squares estimation. Following the working model suggested by Bell
and McCaffrey (2002), in which \(\boldsymbol\Phi = \mathbf{I}\), the
theorem shows that the adjustment method is invariant to the choice of
estimator so long as the model is estimated by OLS (i.e., with
\(\mathbf{W} = \mathbf{I}\)). In contrast, if the working model proposed
by Imbens and Kolesar (2016) is instead used (while still using OLS),
then the the CR2 adjustments might differ depending on whether LSDV or
the fixed effects estimator is used.

\hypertarget{sec:testing}{%
\section{HYPOTHESIS TESTING}\label{sec:testing}}

The CR2 correction produces a CRVE that has reduced bias (compared to
other CRVEs) when the number of clusters is small, leading to more
accurate standard errors. However, standard errors are of limited
inherent interest. Rather, their main use is for the construction of
hypothesis tests and confidence intervals, which are often based on
Wald-type test statistics.

Cluster-robust Wald tests are justified on an asymptotic basis as the
number of clusters grows large. Evidence from a wide variety of contexts
indicates that the asymptotic limiting distribution of robust Wald
statistics can be a poor approximation when the number of clusters is
small, even if corrections such as CR2 or CR3 are employed (Bell and
McCaffrey 2002; Bertrand, Duflo, and Mullainathan 2004; Cameron,
Gelbach, and Miller 2008). Like the bias of the CRVE estimator itself,
the accuracy of the asymptotic approximations depends on design features
such as the degree of imbalance across clusters, skewness or leverage of
the covariates, and the similarity of cluster sizes (Tipton and
Pustejovsky 2015; McCaffrey, Bell, and Botts 2001; MacKinnon and Webb
2016; Carter, Schnepel, and Steigerwald 2013). This suggests that, if
hypothesis tests are to achieve accurate rejection rates in small
samples, they should account for features of the design matrix.

In this section, we develop a method for testing linear constraints on
\(\boldsymbol\beta\), where the null hypothesis has the form
\(H_0: \mathbf{C}\boldsymbol\beta = \mathbf{d}\) for fixed
\(q \times r\) matrix \(\mathbf{C}\) and \(q \times 1\) vector
\(\mathbf{d}\). The cluster-robust Wald statistic is then
\begin{equation}
\label{eq:Wald_stat}
Q = \left(\mathbf{C}\boldsymbol{\hat\beta} - \mathbf{d}\right)'\left(\mathbf{C} \mathbf{V}^{CR} \mathbf{C}'\right)^{-1}\left(\mathbf{C}\boldsymbol{\hat\beta} - \mathbf{d}\right),
\end{equation} where \(\mathbf{V}^{CR}\) is one of the cluster-robust
estimators described in previous sections. The asymptotic Wald test
rejects \(H_0\) if \(Q\) exceeds the \(\alpha\) critical value from a
chi-squared distribution with \(q\) degrees of freedom. It can be shown
that this test approaches level \(\alpha\) when the number of clusters
is large. However, in practice it is rarely clear how large a sample is
needed for the asymptotic approximation to be accurate.

\hypertarget{subsec:t-tests}{%
\subsection{Small-sample corrections for t-tests}\label{subsec:t-tests}}

Consider testing the hypothesis \(H_0: \mathbf{c}'\boldsymbol\beta = 0\)
for a fixed \(r \times 1\) contrast vector \(\mathbf{c}\). For this
one-dimensional constraint, an equivalent to the Wald statistic given in
(\ref{eq:Wald_stat}) is to use the test statistic
\(Z = \mathbf{c}'\boldsymbol{\hat\beta} / \sqrt{\mathbf{c}'\mathbf{V}^{CR}\mathbf{c}}\),
which follows a standard normal distribution in large samples. In small
samples, it is common to use the CR1 or CR1S estimator and to
approximate the distribution of \(Z\) by a \(t(m - 1)\) distribution.
Hansen (2007) provided one justification for the use of this reference
distribution by identifying conditions under which \(Z\) converges in
distribution to \(t(m-1)\) as the within-cluster sample sizes grow
large, with \(m\) fixed (see also Donald and Lang 2007). Ibragimov and
Müller (2010) proposed a weighting technique derived so that \(t(m-1)\)
critical values lead to rejection rates less than or equal to
\(\alpha\). Both of these arguments require that
\(\mathbf{c}'\boldsymbol\beta\) be separately identified within each
cluster. Outside of these circumstances, using \(t(m-1)\) critical
values can still lead to over-rejection (Cameron and Miller 2015).
Furthermore, using these critical values does not take into account that
the distribution of \(\mathbf{V}^{CR}\) is affected by the structure of
\(\mathbf{X}\).

Bell and McCaffrey (2002) proposed to compare \(Z\) to a \(t(\nu)\)
references distribution, with degrees of freedom \(\nu\) estimated by a
Satterthwaite approximation. The Satterthwaite approximation
(Satterthwaite 1946) entails using degrees of freedom that are a
function of the the first two moments of the sampling distribution of
\(\mathbf{c}' \mathbf{V}^{CR} \mathbf{c}\). Expressions for the first
two moments of \(\mathbf{c}'\mathbf{V}^{CR2}\mathbf{c}\) can be derived
under the assumption that the errors
\(\boldsymbol\epsilon_1,...,\boldsymbol\epsilon_m\) are normally
distributed. In practice, both moments involve the variance structure
\(\boldsymbol\Sigma\), which is unknown. Bell and McCaffrey (2002)
proposed to estimate the moments based on the same working model that is
used to derive the adjustment matrices. This ``model-assisted'' estimate
of the degrees of freedom is then calculated as \begin{equation}
\label{eq:nu_model}
\nu_{M} = \frac{\left(\sum_{i=1}^m \mathbf{p}_i' \boldsymbol\Phi \mathbf{p}_i\right)^2}{\sum_{i=1}^m \sum_{j=1}^m \left(\mathbf{p}_i' \boldsymbol\Phi \mathbf{p}_j\right)^2},
\end{equation} where
\(\mathbf{p}_i = \left(\mathbf{I} - \mathbf{H_X}\right)_i'\mathbf{A}_i \mathbf{W}_i\mathbf{\ddot{R}}_i\mathbf{M_{\ddot{R}}} \mathbf{c}\).
This approximation works because the degrees of freedom account for
covariate features that affect the distribution of the test statistic.

Previous simulation studies have examined the performance of t-tests
based on the CR2 variance estimator and Satterthwaite approximation
under a variety of conditions, including panel data models (Cameron and
Miller 2015; Imbens and Kolesar 2016), analysis of multi-stage surveys
(Bell and McCaffrey 2002), and meta-analysis (Tipton 2015). Across this
range of data-generating processes, these studies found that the Type I
error rate of the test is nearly always less than or equal to the
nominal \(\alpha\), so long as the degrees of freedom are larger than 4
or 5 (Tipton 2015; Bell and McCaffrey 2002). Because the degrees of
freedom are covariate-dependent, it is not possible to assess whether a
small-sample correction is needed based solely on the total number of
clusters in the data. Consequently, Tipton (2015) and Imbens and Kolesar
(2016) argued that t-tests based on CRVE should routinely use the CR2
variance estimator and the Satterthwaite degrees of freedom, even when
\(m\) appears to be large.

\hypertarget{subsec:F-tests}{%
\subsection{Small-sample corrections for F-tests}\label{subsec:F-tests}}

Little research has considered small-sample corrections for
multiple-constraint hypothesis tests based on cluster-robust Wald
statistics. Cameron and Miller highlighted this problem, noting that
some form of adjustment is clearly needed in light of the extensive work
on single-parameter tests. We now describe an approach to
multi-parameter testing that closely parallels the Satterthwaite
correction for t-tests.

Our approach is to approximate the sampling distribution of \(Q\) by
Hotelling's \(T^2\) distribution (a multiple of an F distribution) with
estimated degrees of freedom. To motivate the approximation, let
\(\mathbf{G} = \mathbf{C} \mathbf{M_{\ddot{R}}}\mathbf{\ddot{R}}'\mathbf{W}\boldsymbol\Phi\mathbf{W}\mathbf{\ddot{R}}\mathbf{M_{\ddot{R}}} \mathbf{C}'\)
denote the variance of \(\mathbf{C}\boldsymbol{\hat\beta}\) under the
working model and observe that \(Q\) can be written as
\(Q = \mathbf{z}' \boldsymbol\Omega^{-1} \mathbf{z}\), where
\(\mathbf{z} = \mathbf{G}^{-1/2}\left(\mathbf{C}\boldsymbol{\hat\beta} - \mathbf{d}\right)\)
and
\(\boldsymbol\Omega = \mathbf{G}^{-1/2} \mathbf{C} \mathbf{V}^{CR}\mathbf{C}'\mathbf{G}^{-1/2}\).
Now suppose that \(\eta \times \boldsymbol\Omega\) follows a Wishart
distribution with \(\eta\) degrees of freedom and a \(q\)-dimensional
identity scale matrix. It then follows that \begin{equation}
\label{eq:AHT}
\left(\frac{\eta - q + 1}{\eta q}\right) Q \ \dot\sim \ F(q, \eta - q + 1).
\end{equation} We will refer to this as the approximate Hotelling's
\(T^2\) (AHT) test. We consider how to estimate \(\eta\) below. Note
that this approximation reduces to the Satterthwaite approximation when
\(q = 1\). For \(q > 1\), the test depends on the multivariate
distribution of \(\mathbf{V}^{CR}\), including both variance and
covariance terms.

Tipton and Pustejovsky (2015) recently introduced this test for the
special case of CRVE for regression models used in meta-analysis.
Wishart approximations have also been considered as approximations in
several simpler models where special cases of CRVEs are used. Nel and
Merwe (1986) proposed an AHT-type test for equality of multivariate
means across two samples with unequal variance-covariance matrices
(i.e., the multivariate Behrens-Fisher problem, see also Krishnamoorthy
and Yu 2004). Zhang (2012) followed a similar approach in developing a
test for contrasts in multivariate analysis of variance models where the
covariance of the errors differs across groups, a special case of model
(\ref{eq:fixed_effects}) where the CR2 variance estimator has a
particularly simple form. In each of these special cases, the robust
variance estimator is a mixture of Wishart distributions that is
well-approximated by a Wishart distribution with estimated degrees of
freedom. Additionally, Pan and Wall (2002) described an F-test for use
in GEE models, which uses the Wishart approximation to the distribution
of \(\mathbf{V}^{CR0}\) but estimates the degrees of freedom using a
different method than the one we describe below.

In an extensive simulation, Tipton and Pustejovsky (2015) compared the
performance of the AHT test to several other possible approximate
F-tests, including adaptations of the tests introduced by Pan and Wall
(2002) and Zhang (2012), as well as adaptations of tests based on
eigen-decompositions proposed by Fai and Cornelius (1996) and Cai and
Hayes (2008). Simulation results indicated that the AHT test presented
here has Type I error closer to nominal than any of the other tests
across a wide range of parameter values, covariate types, and
hypotheses. The contribution of the present paper is to extend the AHT
test to the general setting of linear models with fixed effects and
clustered errors.

The remaining question is how to estimate the parameter \(\eta\), which
determines the scalar multiplier and denominator degrees of freedom of
the AHT test. To do so, we match the mean and variance of
\(\boldsymbol\Omega\) to that of the approximating Wishart distribution
under the working variance model \(\boldsymbol\Phi\), just as in the
Satterthwaite degrees of freedom approximation for the t-test. However,
it is not possible to exactly match both moments if \(q > 1\). Following
Tipton and Pustejovsky (2015), we instead match the mean and total
variance of \(\boldsymbol\Omega\) (i.e., the sum of the variances of its
entries).

Let \(\mathbf{g}_1,...,\mathbf{g}_q\) denote the \(q \times 1\) column
vectors of \(\mathbf{G}^{-1/2}\). Let \[
\mathbf{p}_{si} = \left(\mathbf{I} - \mathbf{H_X}\right)_i' \mathbf{A}_i \mathbf{W}_i \mathbf{\ddot{R}}_i \mathbf{M_{\ddot{R}}}\mathbf{C} \mathbf{g}_s \]
for \(s = 1,...,q\) and \(i = 1,...,m\). Under the working model, the
degrees of freedom are then approximated as \begin{equation}
\label{eq:eta_model}
\eta_M = \frac{q(q + 1)}{\sum_{s,t=1}^q \sum_{i,j=1}^m \left(\mathbf{p}_{si}'\boldsymbol\Phi\mathbf{p}_{tj} \mathbf{p}_{ti}'\boldsymbol\Phi\mathbf{p}_{sj} + \mathbf{p}_{si}'\boldsymbol\Phi\mathbf{p}_{sj} \mathbf{p}_{ti}'\boldsymbol\Phi\mathbf{p}_{tj}\right)}.
\end{equation} If \(q = 1\), then \(\eta_M\) reduces to \(\nu_M\) from
Equation (\ref{eq:nu_model}).

This AHT F-test shares several features with the Satterthwaite
approximation for t-tests. As with the t-test, the degrees of freedom of
this F-test depend not only on the number of clusters, but also on
features of the covariates being tested. The degrees of freedom can be
much lower than \(m - 1\), particularly when the covariates being tested
exhibit high leverage or are unbalanced across clusters. For example, if
the goal is to test if there are differences across a three-arm,
block-randomized experiment with clustering by block, the degrees of
freedom will be largest (approaching \(m - 1\)) when the treatment is
allocated equally across the three groups within each block. If the
treatment allocation varies from cluster to cluster, the degrees of
freedom will be smaller---even if the total number of clusters is large.
We thus expect that using the AHT degrees of freedom, which take into
account features of the covariate distribution, will improve the
accuracy of the rejection rates in small samples.

\hypertarget{sec:simulation}{%
\section{SIMULATION STUDY}\label{sec:simulation}}

Evidence from several large simulation studies indicates that hypothesis
tests based on the CR2 adjustment and estimated degrees of freedom
substantially out-perform the procedures that are most commonly used in
empirical applications (Tipton 2015; Tipton and Pustejovsky 2015;
Cameron and Miller 2015; Imbens and Kolesar 2016; Bell and McCaffrey
2002). However, existing simulations have focused almost entirely on
single-parameter tests. In this section, we describe the design and
results of a new simulation study, which focused on the rejection rates
of multiple-parameter tests. Throughout, we refer to tests employing the
CR2-corrected CRVE and estimated degrees of freedom as ``AHT'' tests;
for t-tests, the estimated degrees of freedom are equivalent to the
Satterthwaite approximation given in Equation (\ref{eq:nu_model}). We
refer to tests employing the CR1 correction and \(m - 1\) degrees of
freedom as ``standard'' tests.

\hypertarget{design}{%
\subsection{Design}\label{design}}

The simulation study examined the performance of hypothesis tests on the
relative effects of three policy conditions, in each of three distinct
study designs. First, we considered a randomized block (RB) design in
which every policy condition is observed in every cluster. Second, we
considered a cluster-randomized (CR) design in which each cluster is
observed under a single policy condition. Third, we considered a
difference-in-differences (DD) design in which some clusters are
observed under all three policy conditions while other clusters are
observed under a single condition. For each design, we simulated both
balanced and unbalanced configurations, with \(m = 15\), 30, or 50
cluster and \(n = 6\), 18, or 30 units per cluster;
\ref{app:simulations} describes the exact specifications used. To induce
further imbalance into the designs, we also manipulated whether the
outcomes were fully observed or were missing for 15\% of observations
(completely at random). Appendix Table \ref{tab:simulation_parameters}
summarizes the design of the simulation.

In order to examine the performance of the proposed testing procedures
for constraints of varying dimension, we simulated tri-variate outcome
data. Letting \(y_{hijk}\) denote the measurement of outcome \(k\) at
time point \(j\) for unit \(i\) under condition \(h\), for
\(h = 1,2,3\), \(i = 1,...,m\), \(j = 1,...,n\), and \(k = 1,2,3\), we
generated data according to \begin{equation}
\label{eq:data_generating_model}
y_{hijk} = \mu_{i} + \delta_{hi} + \epsilon_{ijk},
\end{equation} where \(\mu_i\) is the mean outcome for unit \(i\);
\(\delta_{hi}\) is a random treatment effect for unit \(i\) under
condition \(h\), with \(\delta_{1i} = 0\); and \(\epsilon_{ijk}\) is the
idiosyncratic error for unit \(i\) at time point \(j\) on outcome \(k\).
The means \(\mu_1,...,\mu_m\) were sampled from a normal distribution
with mean 0 and variance \(\tau^2\). The unit-specific treatment effects
\(\delta_{2i},\delta_{3i}\) were taken to follow a bivariate normal
distribution with mean zero,
\(\text{Var}\left(\delta_{hi}\right) = \sigma_\delta^2\) for
\(h = 2,3\), and
\(\text{corr}\left(\delta_{2i},\delta_{3i}\right) = 0.9\).\\
The errors at a given time point were assumed to be correlated, with
\(\text{Var}\left(\epsilon_{ijk}\right) = 1 - \tau^2\) and
\(\text{corr}\left(\epsilon_{ijk}, \epsilon_{ijl}\right) = \rho\) for
\(k\neq l\), \(k,l = 1,2,3\). Under this data-generating process, we
simulated data with intra-class correlations of \(\tau^2 = 0.05\), 0.15,
or 0.25; treatment effect variability of \(\sigma_\delta^2 = 0.00\),
0.04, or 0.09; and outcomes that were either weakly (\(\rho = .2\)) or
strongly correlated (\(\rho = .8\)). In combination with the study
design factors, the simulation therefore included a total of 1944 unique
conditions.

Given a set of simulated data, we estimated the effects of the second
and third policy conditions (relative to the first) on each outcome
using a seemingly unrelated regression. For the
difference-in-differences design, we used the analytic model
\begin{equation}
\label{eq:sim_analytic_model}
y_{hijk} = \beta_{hk} + \mu_i + \gamma_j + \epsilon_{ijk},
\end{equation} where \(\beta_{hk}\) is the mean of outcome \(k\) under
condition \(h\), \(\mu_i\) is a fixed effect for each cluster,
\(\gamma_j\) is a fixed effect for each unit within the cluster (i.e.,
per time-point), and \(\epsilon_{ijk}\) is residual error. For the
cluster-randomized designs, fixed effects for clusters were omitted
because the clusters are nested within treatment conditions. For the
randomized block designs, treatments were blocked by cluster and the
fixed effects for time-points were omitted for simplicity. The analytic
model was estimated by OLS after absorbing any fixed effects, and so the
``working'' model amounts to assuming that the errors are independent
and identically distributed. Note that the true data generating model
departs from the working model because of correlation among the outcomes
(\(\rho > 0\)) and because of treatment effect variability
(\(\sigma_\delta^2 > 0\)); in the cluster-randomized designs, working
model misspecification also arises from the intra-class correlation
(\(\tau^2 > 0\)). The range of parameter combinations used in the true
data generating model thus allows us to examine the performance of AHT
tests under varying degrees of working model misspecification.

We tested several single- and multi-parameter constraints on analytic
model (\ref{eq:sim_analytic_model}). We first tested the
single-dimensional null hypotheses that a given policy condition had no
average effect on the first outcome (\(H_0: \mu_{11} = \mu_{12}\) or
\(H_0: \mu_{11} = \mu_{13}\)). We also tested the null hypothesis of no
differences among policy conditions on the first outcome
(\(H_0: \mu_{11} = \mu_{12} = \mu_{13}\)), which has dimension
\(q = 2\). We then tested the multi-variate versions of the above tests,
which involve all three outcome measures jointly and so have dimension
\(q = 3\) or \(q = 6\). For a given combination of study design, sample
sizes, and parameter values, we simulated 50,000 datasets from model
(\ref{eq:data_generating_model}), estimated model
(\ref{eq:sim_analytic_model}) on each dataset, and computed all of the
hypothesis tests. Simulated Type I error rates therefore have standard
errors of approximately 0.0010 for \(\alpha = .05\).

\hypertarget{simulation-results}{%
\subsection{Simulation Results}\label{simulation-results}}

\begin{sidewaysfigure}

{\centering \includegraphics[width=\linewidth]{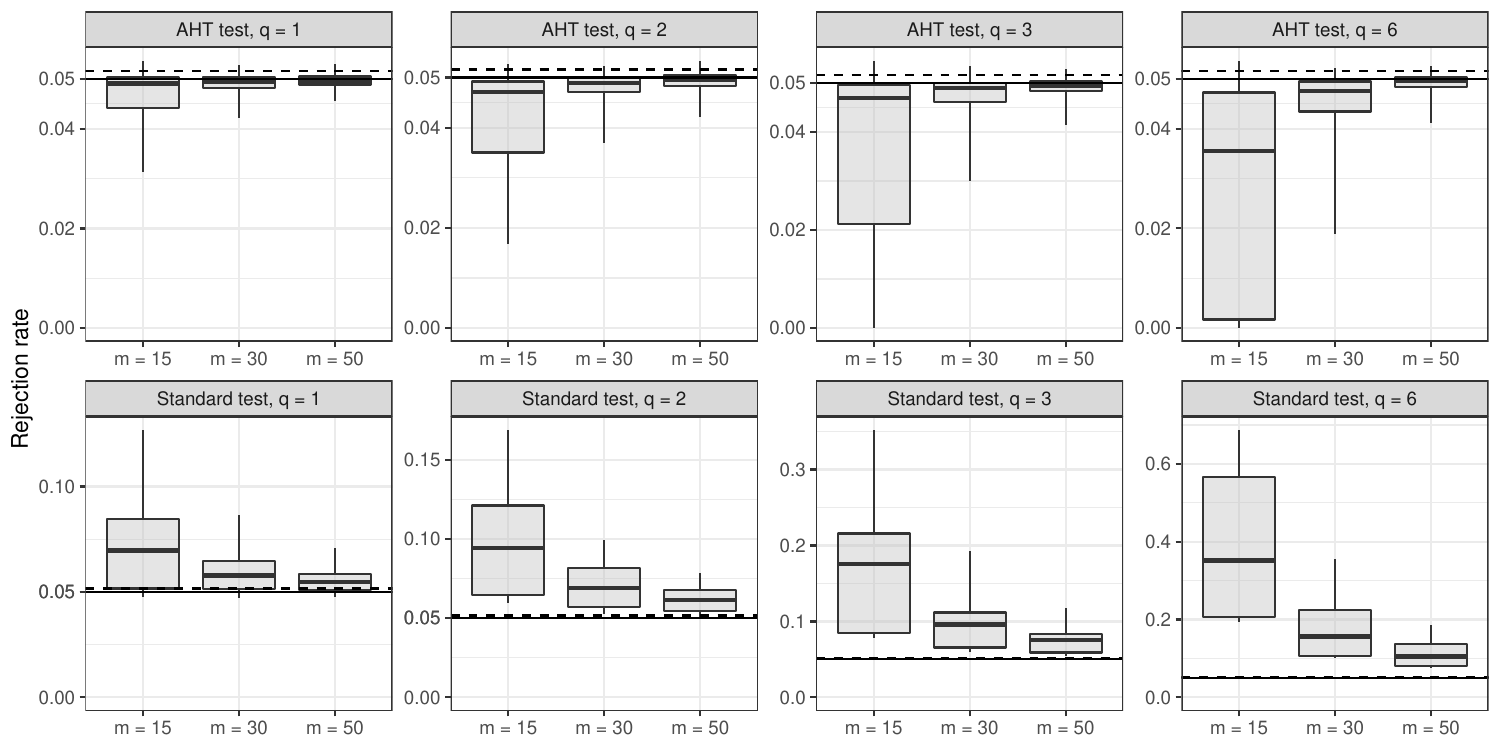} 

}

\caption{Rejection rates of AHT and standard tests for $\alpha = .05$, by dimension of hypothesis ($q$) and sample size ($m$). The solid horizontal line indicates the stated $\alpha$ level and the dashed line indicates an upper confidence bound on simulation error.}\label{fig:overview}
\end{sidewaysfigure}

We discuss five main findings from the simulation results. Throughout,
we present results for the nominal Type I error rate of
\(\alpha = .05\); results for other \(\alpha\) levels are available in
\ref{app:sim-results}. Supplementary materials include complete
numerical results and replication code.

\hypertarget{overall-performance.}{%
\paragraph{Overall performance.}\label{overall-performance.}}

The first finding is that the AHT test has Type I error close to the
stated \(\alpha\) level for all parameter combinations studied, whereas
the standard test based on CR1 does not. Figure \ref{fig:overview}
illustrates this pattern at the nominal type I error rate of
\(\alpha = .05\), for constraints of varying dimension (from \(q = 1\),
in the first column, to \(q = 6\), in the final column) and varying
number of clusters. It can be seen that the AHT test nearly always has
Type I error less than or equal to the stated \(\alpha\) level, even
with a small number of clusters. When the number of clusters is very
small, the Type I error can be smaller than \(\alpha\), particularly for
tests of multiple-constraint hypotheses. Although the error is above the
simulation bound under some conditions, the departures are typically
small. For example, when \(m = 15\) the rejection rates do not exceed
0.012 for \(\alpha = .01\), 0.055 for \(\alpha = .05\), and 0.106 for
\(\alpha = .10\). The rejection rates are even closer to nominal for
lower-dimensional constraints.

In comparison to the AHT test, the Type I error for the standard test
can be markedly higher than the stated \(\alpha\) level, particularly
when the number of clusters is small or the dimension of the hypothesis
is large. For example, the maximum Type I error ranges from 0.127
(\(q = 1\)) to 0.687 (\(q = 6\)) for data sets with 15 clusters. Perhaps
even more important for practice, the rejection rate of the standard
test can be far above the stated \(\alpha\) level even when there are 50
clusters, with maximum error ranging from 0.071 (\(q = 1\)) to 0.186
(\(q = 6\)).

\begin{sidewaysfigure}

{\centering \includegraphics[width=\linewidth]{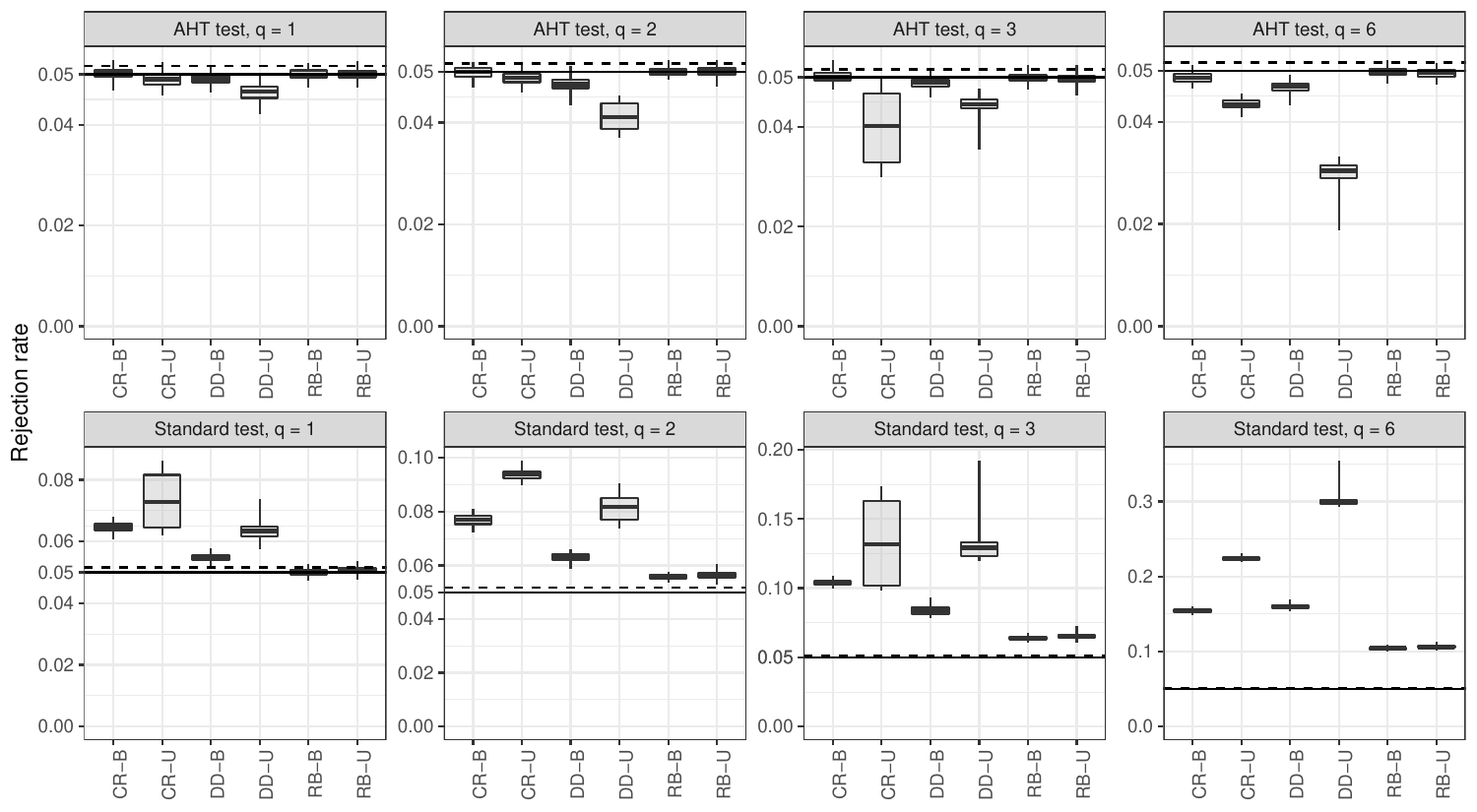} 

}

\caption{Rejection rates of AHT and standard tests, by study design and dimension of hypothesis ($q$) for $\alpha = .05$ and $m = 30$. The solid horizontal line indicates the stated $\alpha$ level and the dashed line indicates an upper confidence bound on simulation error. CR = cluster-randomized design; DD = difference-in-differences design; RB = randomized block design; B = balanced; U = unbalanced.}\label{fig:balance}
\end{sidewaysfigure}

\hypertarget{role-of-balance.}{%
\paragraph{Role of balance.}\label{role-of-balance.}}

Figure \ref{fig:balance} breaks out the rejection rates by study design,
focusing on a sample size of \(m = 30\). In the bottom row, it can be
seen that the rejection rate of the standard test increases with the
dimension of the test (\(q\)) and the degree of unbalance in the study
design. Differences between the balanced and unbalanced designs are
largest for the CR and DD designs, with smaller discrepancies in RB
designs. In the top row, rejection rates of the AHT test are near or
below nominal (between 0.019 and 0.054) across all conditions with at
least 30 clusters. Unbalanced designs lead to rejection rates that are
usually below the nominal \(\alpha\)---just the opposite of how the
standard test is affected by unbalance. This trend is the strongest for
CR and DD designs. At the smallest sample size, the rejection rates of
the AHT test are close to 0 in unbalanced CR and DD designs (see
Appendix Figures \ref{fig:balance_005_15}-\ref{fig:balance_10_15}).

\begin{figure}

{\centering \includegraphics[width=\linewidth]{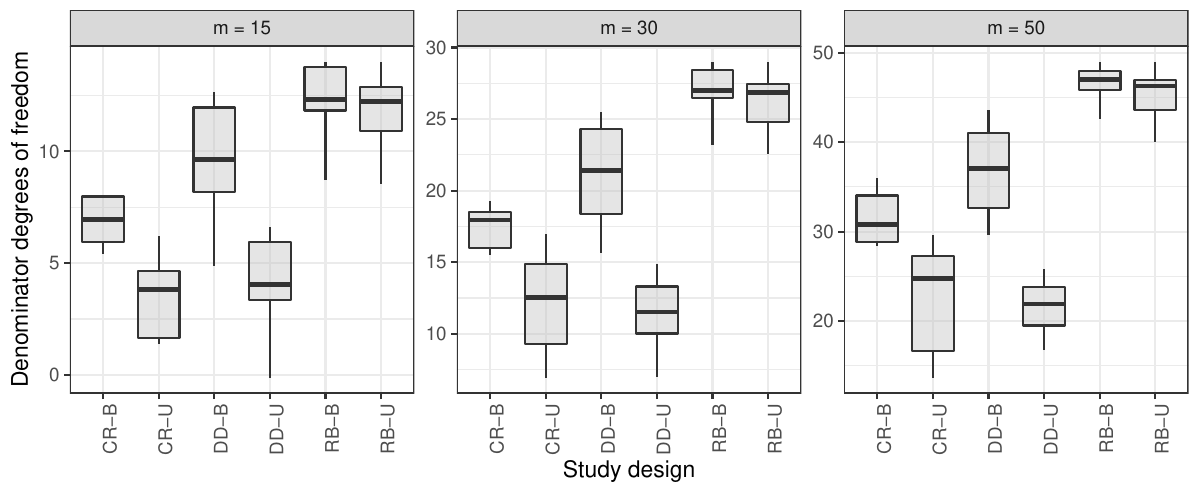} 

}

\caption{Range of denominator degrees of freedom for AHT test, by number of clusters and study design. CR = cluster-randomized design; DD = difference-in-differences design; RB = randomized block design; B = balanced; U = unbalanced.}\label{fig:df}
\end{figure}

\hypertarget{degrees-of-freedom.}{%
\paragraph{Degrees of freedom.}\label{degrees-of-freedom.}}

Figure \ref{fig:df} depicts the range of estimated degrees of freedom
for the AHT test as a function of the study design and number of
clusters (\(m\)). Within each box plot, the degrees of freedom vary
depending on the hypothesis tested, with constraints of larger dimension
having lower degrees of freedom. It can be seen that the AHT degrees of
freedom are often far less than \(m - 1\) and that they are strongly
affected by the pattern of treatment assignment and the degree of
balance. The balanced and unbalanced RB designs have AHT degrees of
freedom closest to \(m - 1\) because the treatment effects being tested
are all identified within each cluster. The balanced DD design usually
has the next largest degrees of freedom because it involves contrasts
between two patterns of treatment configuration, followed by the
balanced CR design, which involves contrasts between three patterns of
treatment configurations. For both of these designs, unbalance leads to
sharply reduced degrees of freedom.

\hypertarget{other-correction-factors.}{%
\paragraph{Other correction factors.}\label{other-correction-factors.}}

In principle, the AHT test could be applied with other forms of CRVEs,
or could be calculated without accounting for absorption of the fixed
effects. Appendix Figures \ref{fig:absorption_005} through
\ref{fig:absorption_10} in the supplementary materials depict the
rejection rates of two variants of the AHT test, including the test
based on the CR1 adjustment matrices and the test based on CR2, but
without accounting for absorption of fixed effects. Using the AHT
degrees of freedom with the CR1 variance estimator leads to a drastic
improvement compared to the standard test with \(m - 1\) degrees of
freedom, although the rejection rates still exceed the nominal level
under some conditions. The test based on CR2 without accounting for
absorption has quite accurate rejection rates, although they exceed
nominal levels more often than those of the test that does account for
absorption, particularly for lower-dimensional hypotheses.

\hypertarget{misspecification.}{%
\paragraph{Misspecification.}\label{misspecification.}}

By simulating the errors across a variety of parameter combinations, we
were also able to test the impact of misspecification of the working
model on Type I error. Because the CR2 correction and AHT degrees of
freedom are both based on a working model with independent,
homoskedastic errors, model misspecification increases with the true
level of treatment effect variance (\(\sigma_\delta^2\)) and intra-class
correlation (\(\tau^2\)). Across the nine error structures, the range of
rejection rates remains very similar for the AHT test, with no clear
pattern related to the degree of mis-specification (see Appendix Figures
\ref{fig:misspecification_005}-\ref{fig:misspecification_10}). These
findings follow closely those from Tipton and Pustejovsky (2015), which
indicated that the Type I error of the AHT test was close to nominal
even when the working model was quite discrepant from the
data-generating model.

In summary, these results demonstrate that the standard robust Wald
test, using the CR1 correction and \(m - 1\) degrees of freedom,
produces a wide range of rejection rates, often far in excess of the
nominal Type I error. In contrast, the rejection rates of the AHT tests
are below or at most slightly above nominal, across all of the
conditions examined. This is because the AHT test incorporates
information about the covariate features into its estimated degrees of
freedom, whereas the standard test does not. Compared to using the AHT
test with the CR1 variance estimator or the modified CR2 estimator that
ignores absorption of fixed effects, the AHT test based on CR2 leads to
rejection rates that are closer to maintaining the nominal level,
although the differences are often fairly small. An important question
that remains is how much the standard test and AHT test diverge in
actual application.

\hypertarget{sec:examples}{%
\section{EXAMPLES}\label{sec:examples}}

This section presents two examples that illustrate the performance of
CRVE in different contexts. In the first example, the effects of
substantive interest involve between-cluster contrasts. The second
example involves a cluster-robust Hausman test for differences between
within- and across-cluster information. In each example, we demonstrate
the proposed AHT test for single- and multiple-parameter hypotheses and
compare the results to the standard test based on the CR1 variance
estimator and \(m - 1\) degrees of freedom. The focus here is on
providing insight into the conditions under which the methods diverge in
terms of three quantities of interest: the standard error estimates, the
degrees of freedom estimates, and the stated p-values. Data files and
replication code are available in the supplementary materials.

\hypertarget{achievement-awards-demonstration}{%
\subsection{Achievement Awards
demonstration}\label{achievement-awards-demonstration}}

Angrist and Lavy (2009) reported results from a randomized trial in
Israel that aimed to increase completion rates of the Bagrut, the
national matriculation certificate for post-secondary education, among
low-achieving high school students. In the Achievement Awards
demonstration, 40 non-vocational high schools with low rates of Bagrut
completion were selected from across Israel, including 10 Arab and 10
Jewish religious schools and 20 Jewish secular schools. The schools were
then pair-matched based on 1999 rates of Bagrut completion, and within
each pair one school was randomized to receive a cash-transfer program.
In these treatment schools, seniors who completed certification were
eligible for payments of approximately \$1,500. Student-level covariate
and outcome data were drawn from administrative records for the school
years ending in June of 2000, 2001, and 2002. The incentive program was
in effect for the group of seniors in treatment schools taking the
Bagrut exams in Spring of 2001, but the program was discontinued for the
following year. We therefore treat completion rates for 2000 and 2002 as
being unaffected by treatment assignment. The primary outcome of
interest is Bagrut completion.

This study involves relatively few clusters, with treatment assigned at
the cluster level. For simplicity, we restrict our analysis to the
sample of female students, which reduces the total sample to 35 schools.
Following the original analysis of Angrist and Lavy (2009), we allow the
program's effects to vary depending on whether a students was in the
upper or lower half of the distribution of prior-year academic
performance. Letting \(h = 1,2,3\) index the sector of each school (Arab
religious, Jewish religious, or Jewish secular), we consider the
following analytic model: \begin{equation}
\label{eq:AL_ATE}
y_{hitj} = z_{hit}\mathbf{r}_{hitj}'\boldsymbol\beta_h + \mathbf{s}_{hitj}'\boldsymbol\gamma + \gamma_{ht} + \mu_{i} + \epsilon_{hitj}
\end{equation} In this model for student \(j\) in year \(t\) in school
\(i\) in sector \(h\), \(z_{hit}\) is an indicator equal to one in the
treatment schools for the 2001 school year and otherwise equal to zero;
\(\mathbf{r}_{hitj}\) is a vector of indicators for whether the student
is in the lower or upper half of the distribution of prior academic
performance; and
\(\boldsymbol\beta_h = \left(\beta_{1h}, \beta_{2h}\right)\) is a vector
of average treatment effects for schools in sector \(h\). The vector
\(\mathbf{s}_{hitj}\) includes the following individual student
demographic measures: mother's and father's education, immigration
status, number of siblings, and indicators for each quartile in the
distribution of prior-year academic performance. The model also includes
fixed effects for each sector in each year (\(\gamma_{ht}\)) and for
each school (\(\mu_{i}\)).

Based on Model (\ref{eq:AL_ATE}), we test four hypotheses. First, we
assume that the program effects are constant across sector (i.e.,
\(\boldsymbol\beta_1 = \boldsymbol\beta_2 = \boldsymbol\beta_3 = \boldsymbol\beta\))
and test for whether the program affected completion rates for students
in the upper half of the prior achievement distribution
(\(H_0: \beta_2 = 0\), with \(q = 1\)). Second, we test for whether the
program was effective in either half of the prior academic performance
(\(H_0: \boldsymbol\beta = 0\), with \(q = 2\)), still assuming that
program effects are constant across sector. Third, we test for whether
program effects in the upper half of the prior achievement distribution
are moderated by school sector
(\(H_0: \beta_{21} = \beta_{22} = \beta_{23}\), with \(q = 3\)).
Finally, we conduct a joint test for whether program effects in either
half of the prior achievement distribution are moderated by school
sector
(\(H_0: \boldsymbol\beta_1 = \boldsymbol\beta_2 = \boldsymbol\beta_3\),
with \(q = 4\)).

\begin{table}[bth]
\centering
\caption{Tests of treatment effects in the Achievement Awards Demonstration} 
\label{tab:AAD}
\begin{tabular}{lcrrr}
  \toprule
Hypothesis & Test & F & df & p \\ 
  \midrule
ATE - upper half (q = 1) & Standard & 5.746 & 34.00 & 0.02217 \\ 
   & AHT & 5.169 & 18.13 & 0.03539 \\ 
  ATE - joint (q = 2) & Standard & 3.848 & 34.00 & 0.03116 \\ 
   & AHT & 3.389 & 16.97 & 0.05775 \\ 
   \midrule
Moderation - upper half (q = 2) & Standard & 3.186 & 34.00 & 0.05393 \\ 
   & AHT & 1.665 & 7.84 & 0.24959 \\ 
  Moderation - joint (q = 4) & Standard & 8.213 & 34.00 & 0.00010 \\ 
   & AHT & 3.091 & 3.69 & 0.16057 \\ 
   \bottomrule
\end{tabular}
\end{table}

Table \ref{tab:AAD} reports the results of all four hypothesis tests.
For the first two hypotheses, the AHT test statistics are slightly
smaller than their standard counterparts and the degrees of freedom are
considerably smaller. These differences in degrees of freedom arise
because the treatment was assigned at the cluster level, while the
subgroups varied within each cluster. In contrast, the AHT and standard
tests diverge markedly for the third and fourth hypotheses tests, which
compare treatment effects across sectors and subgroups. For these cases,
the AHT test statistic and degrees of freedom are both considerably
smaller than those from the standard test. This reflects the degree of
unbalance in allocations across sectors (19 Jewish secular, 7 Jewish
religious, and 9 Arab religious schools), combined with cluster-level
randomization. In combination, these smaller test statistics and degrees
of freedom result in larger p-values for the AHT test when compared to
the standard test.

\hypertarget{effects-of-minimum-legal-drinking-age-on-mortality}{%
\subsection{Effects of minimum legal drinking age on
mortality}\label{effects-of-minimum-legal-drinking-age-on-mortality}}

As a second illustration, we draw on a panel data analysis described in
Angrist and Pischke (2014; see also Carpenter and Dobkin 2011). Based on
data from the Fatal Accident Reporting System maintained by the National
Highway Traffic Safety Administration, we estimate the effects of
changes in the minimum legal drinking age over the time period of
1970-1983 on state-level death rates resulting from motor vehicle
crashes. A standard difference-in-differences specification for such a
state-by-year panel is \begin{equation}
\label{eq:MLDA}
y_{it} = \mathbf{r}_{it}'\boldsymbol\beta + \gamma_t + \mu_i + \epsilon_{it}.
\end{equation} In this model, time-point \(t\) is nested within state
\(i\); the outcome \(y_{it}\) is the number of deaths in motor vehicle
crashes (per 100,000 residents) in state \(i\) at time \(t\);
\(\mathbf{r}_{it}\) is a vector of covariates; \(\gamma_t\) is a fixed
effect for time point \(t\); and \(\mu_i\) is an effect for state \(i\).
The vector \(\mathbf{r}_{it}\) consists of a measure of the proportion
of the population between the ages of 18 and 20 years who can legally
drink alcohol and a measure of the beer taxation rate, both of which
vary across states and across time.

We apply both random effects (RE) and fixed effects (FE) approaches to
estimate the effect of lowering the legal drinking age. For the RE
estimates, we use feasible GLS based the assumption that
\(\mu_1,...,\mu_m\) are mutually independent, normally distributed, and
independent of \(\epsilon_{it}\) and \(\mathbf{r}_{it}\). We also report
an artificial Hausman test (Arellano 1993; Wooldridge 2010) for
correlation between the covariates \(\mathbf{r}_{it}\) and the state
effects \(\mu_i\). Such correlation creates bias in the RE estimator of
the policy effect, thus necessitating the use of the FE estimator. The
artificial Hausman test amends model (\ref{eq:MLDA}) to include
within-cluster deviations for the variables of interest, so that the
specification becomes \begin{equation}
y_{it} = \mathbf{r}_{it}\boldsymbol\beta + \mathbf{\ddot{r}}_{it}\boldsymbol\delta + \gamma_t + \mu_i + \epsilon_{it},
\end{equation} where \(\mathbf{\ddot{r}}_{it}\) denotes the
within-cluster deviations of the covariate. The parameter
\(\boldsymbol\delta\) captures the difference between the
between-cluster and within-cluster estimates of \(\boldsymbol\beta\).
With this setup, the artificial Hausman test amounts to testing the null
hypothesis that \(\boldsymbol\delta = \mathbf{0}\), where
\(\boldsymbol\delta\) is estimated using RE.

\begin{table}[bth]
\centering
\caption{Tests of effects of minimum legal drink age and Hausman specification test} 
\label{tab:MLDA}
\begin{tabular}{lcrrr}
  \toprule
Hypothesis & Test & F & df & p \\ 
  \midrule
Random effects & Standard & 8.261 & 49.00 & 0.00598 \\ 
   & AHT & 7.785 & 26.69 & 0.00960 \\ 
  Fixed effects & Standard & 9.660 & 49.00 & 0.00313 \\ 
   & AHT & 9.116 & 24.58 & 0.00583 \\ 
   \midrule
Hausman test & Standard & 2.930 & 49.00 & 0.06283 \\ 
   & AHT & 2.560 & 11.91 & 0.11886 \\ 
   \bottomrule
\end{tabular}
\end{table}

Table \ref{tab:MLDA} displays the results of the tests for the policy
variable and the Hausman tests for each model specification. The results
of the policy effect tests are quite similar across specifications and
versions of the test. Of note is that, for both the RE and FE estimates,
the AHT tests have only half the degrees of freedom of the corresponding
standard tests. For the artificial Hausman test, the AHT test has fewer
than 12 degrees of freedom, which leads to a much larger p-value
compared to using the standard test based on CR1.

\hypertarget{sec:conclusion}{%
\section{CONCLUSION}\label{sec:conclusion}}

Empirical studies in economics often involve modeling data with a
correlated error structure. In such applications, it is now routine to
use cluster-robust variance estimation, which provides asymptotically
valid standard errors and hypothesis tests without making strong
parametric assumptions about the error structure. However, a growing
body of recent work has drawn attention to the shortcomings of CRVE
methods when the data include only a small or moderate number of
independent clusters (Cameron, Gelbach, and Miller 2008; Cameron and
Miller 2015; Imbens and Kolesar 2016; MacKinnon and Webb 2016). In
particular, Wald tests based on CRVE can have rejection rates far in
excess of the nominal Type I error. This problem is compounded by the
fact that the performance of standard Wald tests depends on features of
the study design beyond just the total number of clusters, which can
make it difficult to determine whether standard, asymptotically valid
CRVE methods are accurate.

One promising solution to this problem is to use the bias-reduced
linearization variance estimator (i.e., CR2) proposed by Bell and
McCaffrey (2002), which corrects the CRVE so that it is exactly unbiased
under an analyst-specified working model for the error structure,
together with degrees of freedom estimated based on the same working
model. In this paper, we have extended the CR2 variance estimator so
that it can be applied to test single- or multi-dimensional parameter
constraints in models with fixed effects in multiple dimensions. We join
Imbens and Kolesar (2016) in arguing that the CR2 estimator and
corresponding estimated degrees of freedom for hypothesis tests should
be applied routinely, whenever analysts use CRVE and hypothesis tests
based thereon. Because the performance of standard CRVE methods depends
on features of the study design, the total number of clusters in the
data is an insufficient guide to whether small-sample corrections are
needed. Instead, the clearest way to determine whether small-sample
corrections are needed is simply to calculate them.

The idea of developing small-sample adjustments based on a working model
may seem strange to analysts accustomed to using CRVE---after all, the
whole point of clustering standard errors is to avoid making assumptions
about the error structure. However, simulation studies reported here and
elsewhere (Tipton 2015; Tipton and Pustejovsky 2015) have demonstrated
that the approach is actually robust to a high degree of
misspecification in the working model. Furthermore, while the working
model provides necessary scaffolding when the number of clusters is
small, its influence tends to fall away as the number of clusters
increases, so that the CR2 estimator and AHT maintain the same
asymptotic robustness as standard CRVE methods.

The proposed AHT test involves two adjustments: use of the CR2
adjustment for the variance estimator and use of estimated degrees of
freedom. Our simulation results and empirical examples illustrate that
the degrees of freedom adjustment has a relatively larger influence on
small-sample performance. Even when used with the CR1 adjustment
matrices, the degrees of freedom adjustment leads to much more accurate
rejection rates, although using the CR2 estimator (and accounting for
absorption of fixed effects) appears to be necessary to fully maintain
the nominal level of the test. The approximate degrees of freedom of the
AHT test can be much smaller than the number of clusters, particularly
when the covariates involved in the test involve high leverage or are
unbalanced across clusters. The estimated degrees of freedom are
indicative of the precision of the standard errors, and thus provide
diagnostic information that is similar to the effective sample size
measure proposed by Carter, Schnepel, and Steigerwald (2013). We
therefore recommend that the degrees of freedom be reported along with
standard errors and \(p\)-values whenever the method is applied.

It is interesting to note that the consequences of accounting for fixed
effects estimation with CR2 runs counter to how fixed effects enter into
other small-sample corrections. In particular, including the fixed
effect parameter dimension in the CR1S small sample adjustment can lead
to bias when the per-cluster sample size is small (Cameron and Miller
2015), whereas accounting for the fixed effects with the CR2 estimator
\emph{improves} the accuracy of test rejection rates, although the
difference is comparatively minor. Accounting for fixed effects also
makes hypothesis test results invariant to how the regression is
calculated (whether by LSDV or after absorption), which we see as a
useful feature for ensuring the replicability of one's analysis.

One limitation of our approach is that the rejection rates of the AHT
tests tend to fall below nominal levels when the number of clusters is
very small and the design is unbalanced. The under-rejection can be
severe for tests of multi-dimensional constraints (e.g.,
\(q = 3\)-dimensional hypotheses with \(m = 15\) clusters in an
unbalanced difference-in-differences design). In principle, this problem
could arise either because the proposed degrees of freedom estimator
loses accuracy under these conditions or because the sampling
distribution of the test statistic is no longer well-approximated by
Hotelling's \(T^2\) distribution at all. In previous work, we
investigated several other methods of approximating the degrees of
freedom, but found none that were more accurate than the method describe
in the present paper (Tipton and Pustejovsky 2015). Thus, future work
may need to focus on other approximations to the reference distribution
itself, such as an F distribution in which both the numerator and
denominator degrees of freedom are estimated (cf. Mehrotra 1997) or
computational approximations such as the cluster-wild bootstrap
(MacKinnon and Webb 2016).

Another outstanding limitation of the CR2 variance estimator is that it
is costly (or even infeasible) to compute when the within-cluster sample
sizes are large (MacKinnon 2015). For example, Bertrand, Duflo, and
Mullainathan (2004) analyzed micro-level data from a 21-year panel of
current population survey data, with clustering by state. Their data
included some state-level clusters with over \(n_i = 10,000\) individual
observations. The CR2 adjustment matrices have dimension
\(n_i \times n_i\), and would be very expensive to compute in this
application. Methods for improving the computational efficiency of the
CR2 variance estimator (or alternative estimators that have similar
performance to CR2), should be investigated further.

This paper has developed the CR2 estimator and AHT testing procedure for
weighted least squares estimation of linear regression models.
Extensions to linear regression models with clustering in multiple,
non-nested dimensions (cf. Cameron, Gelbach, and Miller 2011) appear to
be possible, and their utility should be further investigated. McCaffrey
and Bell (2006) have proposed extensions to bias-reduced linearization
for use with generalized estimating equations, and future work should
consider further extensions to other classes of estimators, including
two-stage least squares and generalized method of moments.

\hypertarget{supplementary-materials}{%
\section*{SUPPLEMENTARY MATERIALS}\label{supplementary-materials}}
\addcontentsline{toc}{section}{SUPPLEMENTARY MATERIALS}

Supplementary materials for the simulation study and empirical examples
are available at
\url{https://github.com/jepusto/clubSandwich/tree/main/paper_ClusterRobustTesting/R}.
The materials include R code for the simulation study, complete
numerical results from the simulations, and data and R code for
replicating the empirical examples.

\hypertarget{acknowledgments}{%
\section*{ACKNOWLEDGMENTS}\label{acknowledgments}}
\addcontentsline{toc}{section}{ACKNOWLEDGMENTS}

This article has benefited from the feedback of seminar participants at
the PRIISM Center at New York University, the University of Texas
Population Research Center, and the American Institutes for Research, as
well as an associate editor and referees. The authors thank Dan Knopf
for helpful discussions about the linear algebra behind the
cluster-robust variance estimator, David Figlio and Coady Wing for
advice about empirical applications and context, and Michael Pfaffenmayr
for drawing our attention to an error in an earlier version of Theorem
2.

\newpage

\hypertarget{references}{%
\section*{REFERENCES}\label{references}}
\addcontentsline{toc}{section}{REFERENCES}

\hypertarget{refs}{}
\begin{CSLReferences}{1}{0}
\leavevmode\vadjust pre{\hypertarget{ref-Angrist2009effects}{}}%
Angrist, Joshua D, and Victor Lavy. 2009. {``{The effects of high stakes
high school achievement awards : Evidence from a randomized trial}.''}
\emph{American Economic Review} 99 (4): 1384--414.
\url{https://doi.org/10.1257/aer.99.4.1384}.

\leavevmode\vadjust pre{\hypertarget{ref-Angrist2014mastering}{}}%
Angrist, Joshua D, and Jörn-Steffen Pischke. 2014.
\emph{Mastering'metrics: The Path from Cause to Effect}. Princeton
University Press.

\leavevmode\vadjust pre{\hypertarget{ref-Angrist2009mostly}{}}%
Angrist, Joshua D, and JS Pischke. 2009. \emph{{Mostly harmless
econometrics: An empiricist's companion}}. Princeton, NJ: Princeton
University Press.

\leavevmode\vadjust pre{\hypertarget{ref-Arellano1987computing}{}}%
Arellano, Manuel. 1987. {``{Computing robust standard errors for
within-groups estimators}.''} \emph{Oxford Bulletin of Economics and
Statistics} 49 (4): 431--34.

\leavevmode\vadjust pre{\hypertarget{ref-Arellano1993on}{}}%
---------. 1993. {``{On the testing of correlated effects with panel
data}.''} \emph{Journal of Econometrics} 59 (1-2): 87--97.
\url{https://doi.org/10.1016/0304-4076(93)90040-C}.

\leavevmode\vadjust pre{\hypertarget{ref-Banerjee2014linear}{}}%
Banerjee, Sudipto, and Anindya Roy. 2014. \emph{{Linear Algebra and
Matrix Analysis for Statistics}}. Boca Raton, FL: Taylor \& Francis.

\leavevmode\vadjust pre{\hypertarget{ref-Bell2002bias}{}}%
Bell, Robert M, and Daniel F McCaffrey. 2002. {``{Bias reduction in
standard errors for linear regression with multi-stage samples}.''}
\emph{Survey Methodology} 28 (2): 169--81.

\leavevmode\vadjust pre{\hypertarget{ref-Bertrand2004how}{}}%
Bertrand, Marianne, Esther Duflo, and Sendhil Mullainathan. 2004.
{``{How much should we trust differences-in-differences estimates?}''}
\emph{Quarterly Journal of Economics} 119 (1): 249--75.

\leavevmode\vadjust pre{\hypertarget{ref-Cai2008new}{}}%
Cai, Li, and Andrew F Hayes. 2008. {``{A new test of linear hypotheses
in OLS regression under heteroscedasticity of unknown form}.''}
\emph{Journal of Educational and Behavioral Statistics} 33 (1): 21--40.
\url{https://doi.org/10.3102/1076998607302628}.

\leavevmode\vadjust pre{\hypertarget{ref-Cameron2008bootstrap}{}}%
Cameron, A Colin, Jonah B Gelbach, and Douglas Miller. 2008.
{``{Bootstrap-based improvements for inference with clustered
errors}.''} \emph{The Review of Economics and Statistics} 90 (3):
414--27.

\leavevmode\vadjust pre{\hypertarget{ref-Cameron2011robust}{}}%
Cameron, A Colin, Jonah B Gelbach, and Douglas L Miller. 2011.
{``{Robust inference with multiway clustering}.''} \emph{Journal of
Business {\&} Economic Statistics} 29 (2): 238--49.
\url{https://doi.org/10.1198/jbes.2010.07136}.

\leavevmode\vadjust pre{\hypertarget{ref-Cameron2015practitioners}{}}%
Cameron, A Colin, and Douglas L Miller. 2015. {``{A Practitioner's Guide
to Cluster-Robust Inference}.''} \emph{Journal of Human Resources} 50
(2): 317--72. \url{https://doi.org/10.3368/jhr.50.2.317}.

\leavevmode\vadjust pre{\hypertarget{ref-Carpenter2011minimum}{}}%
Carpenter, Christopher, and Carlos Dobkin. 2011. {``{The minimum legal
drinking age and public health}.''} \emph{Journal of Economic
Perspectives} 25 (2): 133--56.
\url{https://doi.org/10.1257/jep.25.2.133}.

\leavevmode\vadjust pre{\hypertarget{ref-Carter2013asymptotic}{}}%
Carter, Andrew V, Kevin T Schnepel, and Douglas G Steigerwald. 2013.
{``{Asymptotic Behavior of a t Test Robust to Cluster Heterogeneity}.''}

\leavevmode\vadjust pre{\hypertarget{ref-Conley2011inference}{}}%
Conley, Timothy G, and Christopher R Taber. 2011. {``{Inference with
{`Difference in Differences'} with a Small Number of Policy Changes}.''}
\emph{Review of Economics and Statistics} 93 (1): 113--25.

\leavevmode\vadjust pre{\hypertarget{ref-Davis2002estimating}{}}%
Davis, Peter. 2002. {``{Estimating multi-way error components models
with unbalanced data structures}.''} \emph{Journal of Econometrics} 106:
67--95. \url{https://doi.org/10.1016/S0304-4076(01)00087-2}.

\leavevmode\vadjust pre{\hypertarget{ref-Donald2007inference}{}}%
Donald, Stephen G, and Kevin Lang. 2007. {``{Inference with
difference-in-differences and other panel data}.''} \emph{Review of
Economics and Statistics} 89 (2): 221--33.
\url{https://doi.org/10.1162/rest.89.2.221}.

\leavevmode\vadjust pre{\hypertarget{ref-Fai1996approximate}{}}%
Fai, Alex. Hrong-Tai, and PL Cornelius. 1996. {``{Approximate F-tests of
multiple degree of freedom hypotheses in generalized least squares
analyses of unbalanced split-plot experiments}.''} \emph{Journal of
Statistical Computation and Simulation} 54 (4): 363--78.

\leavevmode\vadjust pre{\hypertarget{ref-Hansen2007asymptotic}{}}%
Hansen, Christian B. 2007. {``{Asymptotic properties of a robust
variance matrix estimator for panel data when T is large}.''}
\emph{Journal of Econometrics} 141: 597--620.
\url{https://doi.org/10.1016/j.jeconom.2006.10.009}.

\leavevmode\vadjust pre{\hypertarget{ref-Ibragimov2010tstatistic}{}}%
Ibragimov, Rustam, and Ulrich K Müller. 2010. {``{t-Statistic based
correlation and heterogeneity robust inference}.''} \emph{Journal of
Business \& Economic Statistics} 28 (4): 453--68.
\url{https://doi.org/10.1198/jbes.2009.08046}.

\leavevmode\vadjust pre{\hypertarget{ref-Ibragimov2016inference}{}}%
---------. 2016. {``{Inference with few heterogeneous clusters}.''}
\emph{Review of Economics and Statistics} 98 (1): 83--96.
\url{https://doi.org/10.1162/REST_a_00545}.

\leavevmode\vadjust pre{\hypertarget{ref-Imbens2015robust}{}}%
Imbens, Guido W, and Michal Kolesar. 2016. {``{Robust Standard Errors in
Small Samples: Some Practical Advice}.''} \emph{Review of Economics and
Statistics} forthcoming.

\leavevmode\vadjust pre{\hypertarget{ref-Krishnamoorthy2004modified}{}}%
Krishnamoorthy, K, and Jianqi Yu. 2004. {``{Modified Nel and Van der
Merwe test for the multivariate Behrens--Fisher problem}.''}
\emph{Statistics {\&} Probability Letters} 66 (2): 161--69.
\url{https://doi.org/10.1016/j.spl.2003.10.012}.

\leavevmode\vadjust pre{\hypertarget{ref-Lee2008regression}{}}%
Lee, David S, and David Card. 2008. {``{Regression discontinuity
inference with specification error}.''} \emph{Journal of Econometrics}
142 (2): 655--74. \url{https://doi.org/10.1016/j.jeconom.2007.05.003}.

\leavevmode\vadjust pre{\hypertarget{ref-Liang1986longitudinal}{}}%
Liang, Kung-Yee, and Scott L Zeger. 1986. {``{Longitudinal data analysis
using generalized linear models}.''} \emph{Biometrika} 73 (1): 13--22.

\leavevmode\vadjust pre{\hypertarget{ref-MacKinnon2013thirty}{}}%
MacKinnon, James G. 2013. {``{Thirty years of heteroskedasticity-robust
inference}.''} In \emph{Recent Advances and Future Directions in
Causality, Prediction, and Specification Analysis}, edited by Xiaohong
Chen and Norman R. Swanson. New York, NY: Springer New York.
\url{https://doi.org/10.1007/978-1-4614-1653-1}.

\leavevmode\vadjust pre{\hypertarget{ref-Mackinnon2015wildCIs}{}}%
---------. 2015. {``{Wild cluster bootstrap confidence intervals}.''}
\emph{L'Actualite Economique, Revue d'analyse Economique} 91 (1-2):
11--33.

\leavevmode\vadjust pre{\hypertarget{ref-MacKinnon2016wild}{}}%
MacKinnon, James G, and Matthew D Webb. 2016. {``{Wild bootstrap
inference for wildly different cluster sizes}.''} \emph{Journal of
Applied Econometrics} forthcoming.
\url{https://doi.org/10.1002/jae.2508}.

\leavevmode\vadjust pre{\hypertarget{ref-MacKinnon1985some}{}}%
MacKinnon, James G, and Halbert White. 1985. {``{Some
heteroskedasticity-consistent covariance matrix estimators with improved
finite sample properties}.''} \emph{Journal of Econometrics} 29:
305--25.

\leavevmode\vadjust pre{\hypertarget{ref-Mancl2001covariance}{}}%
Mancl, Lloyd A, and Timothy A DeRouen. 2001. {``{A covariance estimator
for GEE with improved small-sample properties}.''} \emph{Biometrics} 57
(1): 126--34.

\leavevmode\vadjust pre{\hypertarget{ref-McCaffrey2006improved}{}}%
McCaffrey, Daniel F, and Robert M Bell. 2006. {``{Improved hypothesis
testing for coefficients in generalized estimating equations with small
samples of clusters.}''} \emph{Statistics in Medicine} 25 (23):
4081--98. \url{https://doi.org/10.1002/sim.2502}.

\leavevmode\vadjust pre{\hypertarget{ref-McCaffrey2001generalizations}{}}%
McCaffrey, Daniel F, Robert M Bell, and Carsten H Botts. 2001.
{``{Generalizations of biased reduced linearization}.''} In
\emph{Proceedings of the Annual Meeting of the American Statistical
Association}. 1994.

\leavevmode\vadjust pre{\hypertarget{ref-Mehrotra1997improving}{}}%
Mehrotra, Devan V. 1997. {``{Improving the brown-forsythe solution to
the generalized behrens-fisher problem}.''} \emph{Communications in
Statistics - Simulation and Computation} 26 (3): 1139--45.
\url{https://doi.org/10.1080/03610919708813431}.

\leavevmode\vadjust pre{\hypertarget{ref-Nel1986solution}{}}%
Nel, D G, and C A van der Merwe. 1986. {``{A solution to the
multivariate Behrens-Fisher problem}.''} \emph{Communications in
Statistics - Theory and Methods} 15 (12): 3719--35.

\leavevmode\vadjust pre{\hypertarget{ref-Pan2002small}{}}%
Pan, Wei, and Melanie M Wall. 2002. {``{Small-sample adjustments in
using the sandwich variance estimator in generalized estimating
equations.}''} \emph{Statistics in Medicine} 21 (10): 1429--41.
\url{https://doi.org/10.1002/sim.1142}.

\leavevmode\vadjust pre{\hypertarget{ref-Satterthwaite1946approximate}{}}%
Satterthwaite, Franklin E. 1946. {``{An approximate distribution of
estimates of variance components}.''} \emph{Biometrics Bulletin} 2 (6):
110--14.

\leavevmode\vadjust pre{\hypertarget{ref-Tipton2015small-t}{}}%
Tipton, Elizabeth. 2015. {``{Small sample adjustments for robust
variance estimation with meta-regression.}''} \emph{Psychological
Methods} 20 (3): 375--93. \url{https://doi.org/10.1037/met0000011}.

\leavevmode\vadjust pre{\hypertarget{ref-Tipton2015small-F}{}}%
Tipton, Elizabeth, and James E Pustejovsky. 2015. {``{Small-sample
adjustments for tests of moderators and model fit using robust variance
estimation in meta-regression}.''} \emph{Journal of Educational and
Behavioral Statistics} 40 (6): 604--34.

\leavevmode\vadjust pre{\hypertarget{ref-white1984asymptotic}{}}%
White, Halbert. 1984. \emph{Asymptotic Theory for Econometricians}.
Orlando, FL: Academic Press, Inc.

\leavevmode\vadjust pre{\hypertarget{ref-Wooldridge2003cluster}{}}%
Wooldridge, Jeffrey M. 2003. {``{Cluster-sample methods in applied
econometrics}.''} \emph{The American Economic Review} 93 (2): 133--38.

\leavevmode\vadjust pre{\hypertarget{ref-Wooldridge2010econometric}{}}%
---------. 2010. \emph{{Econometric Analysis of Cross Section and Panel
Data}}. 2nd ed. Cambridge, MA: MIT Press.

\leavevmode\vadjust pre{\hypertarget{ref-Young2016improved}{}}%
Young, Alwyn. 2016. {``{Improved, nearly exact, statistical inference
with robust and clustered covariance matrices using effective degrees of
freedom corrections}.''} London School of Economics.
\url{http://personal.lse.ac.uk/YoungA/Improved.pdf}.

\leavevmode\vadjust pre{\hypertarget{ref-Zhang2012twowayANOVA}{}}%
Zhang, Jin-Ting. 2012. {``{An approximate degrees of freedom test for
heteroscedastic two-way ANOVA}.''} \emph{Journal of Statistical Planning
and Inference} 142 (1): 336--46.
\url{https://doi.org/10.1016/j.jspi.2011.07.023}.

\end{CSLReferences}

\newpage

\renewcommand{\thesection}{Appendix \Alph{section}}
\renewcommand{\thesubsection}{\Alph{section}.\arabic{subsection}}
\setcounter{section}{0}
\setcounter{table}{0}
\setcounter{figure}{0}
\setcounter{equation}{0}
\renewcommand{\thefigure}{A\arabic{figure}}
\renewcommand{\thetable}{A\arabic{table}}
\renewcommand{\theequation}{A\arabic{equation}}

\hypertarget{app:proof1}{%
\section{Proof of Theorem 1}\label{app:proof1}}

Consider the matrix \(\mathbf{B}_i\) as defined in
(\ref{eq:CR2_Bmatrix}): \[
\mathbf{B}_i = \mathbf{D}_i\left(\mathbf{I} - \mathbf{H_X}\right)_i \boldsymbol\Phi \left(\mathbf{I} - \mathbf{H_X}\right)_i' \mathbf{D}_i',
\] The Moore-Penrose inverse of \(\mathbf{B}_i\) can be computed from
its eigen-decomposition. Let \(b \leq n_i\) denote the rank of
\(\mathbf{B}_i\). Let \(\boldsymbol\Lambda\) be the \(b \times b\)
diagonal matrix of the positive eigenvalues of \(\mathbf{B}_i\) and
\(\mathbf{V}\) be the \(n_i \times b\) matrix of corresponding
eigen-vectors, so that
\(\mathbf{B}_i = \mathbf{V}\boldsymbol\Lambda\mathbf{V}'\). Then
\(\mathbf{B}_i^+ = \mathbf{V}\boldsymbol\Lambda^{-1}\mathbf{V}'\) and
\(\mathbf{B}_i^{+1/2} = \mathbf{V}\boldsymbol\Lambda^{-1/2}\mathbf{V}'\).
Because the adjustment matrices taken to be
\(\mathbf{A}_i = \mathbf{D}_i' \mathbf{B}_i^{+1/2} \mathbf{D}_i\), we
have that \begin{equation}\begin{aligned}
\label{eq:step1}
\mathbf{\ddot{R}}_i' \mathbf{W}_i \mathbf{A}_i \left(\mathbf{I} - \mathbf{H_X}\right)_i \boldsymbol\Phi \left(\mathbf{I} - \mathbf{H_X}\right)_i' \mathbf{A}_i' \mathbf{W}_i \mathbf{\ddot{R}}_i &= \mathbf{\ddot{R}}_i' \mathbf{W}_i \mathbf{D}_i' \mathbf{B}_i^{+1/2} \mathbf{B}_i \mathbf{B}_i^{+1/2} \mathbf{D}_i \mathbf{W}_i \mathbf{\ddot{R}}_i \\
&= \mathbf{\ddot{R}}_i' \mathbf{W}_i \mathbf{D}_i' \mathbf{V}\mathbf{V}' \mathbf{D}_i \mathbf{W}_i \mathbf{\ddot{R}}_i. 
\end{aligned}\end{equation} Thus, it will suffice to show that
\(\mathbf{V}'\mathbf{D}_i \mathbf{W}_i \mathbf{\ddot{R}}_i = \mathbf{D}_i \mathbf{W}_i \mathbf{\ddot{R}}_i\).

Now, because \(\mathbf{D}_i\) and \(\boldsymbol\Phi\) are positive
definite and \(\mathbf{B}_i\) is symmetric, the eigen-vectors
\(\mathbf{V}\) define an orthonormal basis for the column span of
\(\left(\mathbf{I} - \mathbf{H_X}\right)_i\). We now show that
\(\mathbf{\ddot{U}}_i\) is in the column space of
\(\left(\mathbf{I} - \mathbf{H_X}\right)_i\). Let \(\mathbf{Z}_i\) be an
\(n_i \times (r + s)\) matrix of zeros. With \(\mathbf{L}_i\) as defined
in Theorem \ref{thm:BRL_FE}, take
\(\mathbf{Z}_k = - \mathbf{\ddot{U}}_k \mathbf{L}_i^{-1}\mathbf{M}_{\mathbf{\ddot{U}}}^{-1}\),
for \(k \neq i\) and
\(\mathbf{Z} = \left(\mathbf{Z}_1',...,\mathbf{Z}_m'\right)'\). Observe
that \(\left(\mathbf{I} - \mathbf{H_T}\right) \mathbf{Z} = \mathbf{Z}\).
It follows that \begin{align*}
\left(\mathbf{I} - \mathbf{H_X}\right)_i \mathbf{Z} &= \left(\mathbf{I} - \mathbf{H_{\ddot{U}}}\right)_i \left(\mathbf{I} - \mathbf{H_T}\right) \mathbf{Z} \\
&= \left(\mathbf{I} - \mathbf{H_{\ddot{U}}}\right)_i \mathbf{Z} \\
&= \mathbf{Z}_i - \mathbf{\ddot{U}}_i\mathbf{M_{\ddot{U}}}\sum_{k=1}^m \mathbf{\ddot{U}}_k'\mathbf{W}_k\mathbf{Z}_k \\
&= \mathbf{\ddot{U}}_i\mathbf{M_{\ddot{U}}} \left(\sum_{k \neq i} \mathbf{\ddot{U}}_k' \mathbf{W}_k \mathbf{\ddot{U}} \right) \mathbf{L}_i^{-1}\mathbf{M}_{\mathbf{\ddot{U}}}^{-1} \\
&= \mathbf{\ddot{U}}_i\mathbf{M_{\ddot{U}}} \mathbf{L}_i \mathbf{L}_i^{-1} \mathbf{M}_{\mathbf{\ddot{U}}}^{-1} \\
&= \mathbf{\ddot{U}}_i.
\end{align*} Thus, there exists an \(N \times (r + s)\) matrix
\(\mathbf{Z}\) such that
\(\left(\mathbf{I} - \mathbf{H_{\ddot{X}}}\right)_i \mathbf{Z} = \mathbf{\ddot{U}}_i\),
i.e., \(\mathbf{\ddot{U}}_i\) is in the column span of
\(\left(\mathbf{I} - \mathbf{H_X}\right)_i\). Because
\(\mathbf{D}_i \mathbf{W}_i\) is positive definite and
\(\mathbf{\ddot{R}}_i\) is a sub-matrix of \(\mathbf{\ddot{U}}_i\),
\(\mathbf{D}_i\mathbf{W}_i\mathbf{\ddot{R}}_i\) is also in the column
span of \(\left(\mathbf{I} - \mathbf{H_X}\right)_i\). It follows that
\begin{equation}
\label{eq:step2}
\mathbf{\ddot{R}}_i' \mathbf{W}_i \mathbf{D}_i' \mathbf{V}\mathbf{V}' \mathbf{D}_i \mathbf{W}_i \mathbf{\ddot{R}}_i = \mathbf{\ddot{R}}_i' \mathbf{W}_i \boldsymbol\Phi_i \mathbf{W}_i \mathbf{\ddot{R}}_i.
\end{equation} Substituting (\ref{eq:step2}) into (\ref{eq:step1})
demonstrates that \(\mathbf{A}_i\) satisfies the generalized BRL
criterion (Equation \ref{eq:CR2_criterion} of the main text).

Under the working model, the residuals from cluster \(i\) have mean
\(\mathbf{0}\) and variance \[
\text{Var}\left(\mathbf{\ddot{e}}_i\right) = \left(\mathbf{I} - \mathbf{H_X}\right)_i \boldsymbol\Phi \left(\mathbf{I} - \mathbf{H_X}\right)_i',\]
It follows that \begin{align*}
\text{E}\left(\mathbf{V}^{CR2}\right) &= \mathbf{M_{\ddot{R}}}\left[\sum_{i=1}^m \mathbf{\ddot{R}}_i' \mathbf{W}_i \mathbf{A}_i \left(\mathbf{I} - \mathbf{H_X}\right)_i \boldsymbol\Phi \left(\mathbf{I} - \mathbf{H_X}\right)_i' \mathbf{A}_i \mathbf{W}_i \mathbf{\ddot{R}}_i \right] \mathbf{M_{\ddot{R}}} \\
&= \mathbf{M_{\ddot{R}}}\left[\sum_{i=1}^m \mathbf{\ddot{R}}_i' \mathbf{W}_i \boldsymbol\Phi_i \mathbf{W}_i \mathbf{\ddot{R}}_i \right] \mathbf{M_{\ddot{R}}} \\
&= \text{Var}\left(\boldsymbol{\hat\beta}\right)
\end{align*}

\newpage

\hypertarget{app:proof2}{%
\section{Proof of Theorem 2}\label{app:proof2}}

If \(\mathbf{W}_i = \boldsymbol\Phi_i = \mathbf{I}_i\), then we can
write \(\mathbf{B}_i\) from Equation (\ref{eq:CR2_Bmatrix}) as
\begin{align}
\mathbf{B}_i &= \mathbf{D}_i \left(\mathbf{I} - \mathbf{H_{\ddot{U}}}\right)_i \left(\mathbf{I} - \mathbf{H_T}\right) \boldsymbol\Phi \left(\mathbf{I} - \mathbf{H_T}\right)' \left(\mathbf{I} - \mathbf{H_{\ddot{U}}}\right)_i' \mathbf{D}_i' \nonumber \\ 
&= \left(\mathbf{I} - \mathbf{H_{\ddot{U}}} - \mathbf{H_T}\right)_i \left(\mathbf{I} - \mathbf{H_{\ddot{U}}} - \mathbf{H_T}\right)_i'  \nonumber\\ 
\label{eq:B_i}
&= \left(\mathbf{I}_i - \mathbf{\ddot{U}}_i \mathbf{M_{\ddot{U}}}\mathbf{\ddot{U}}_i' - \mathbf{T}_i \mathbf{M_T}\mathbf{T}_i'\right),
\end{align} where the last equality follows from the fact that
\(\mathbf{\ddot{U}}_i'\mathbf{T}_i = \mathbf{0}\) for \(i = 1,...,m\).
Similarly, we can write \begin{equation}
\label{eq:Btilde_i}
\tilde{\mathbf{B}}_i = \left(\mathbf{I}_i - \mathbf{\ddot{U}}_i \mathbf{M_{\ddot{U}}}\mathbf{\ddot{U}}_i'\right).
\end{equation}

We now show that \(\tilde{\mathbf{A}}_i \mathbf{T}_i = \mathbf{T}_i\).
Denote the rank of \(\mathbf{\ddot{U}}_i\) as
\(u_i \leq \min \left\{n_i, r + s \right\}\) and take the thin QR
decomposition of \(\mathbf{\ddot{U}}_i\) as
\(\mathbf{\ddot{U}}_i = \mathbf{Q}_i \mathbf{R}_i\), where
\(\mathbf{Q}_i\) is an \(n_i \times u_i\) semi-orthonormal matrix and
\(\mathbf{R}_i\) is a \(u_i \times r + s\) matrix of rank \(u_i\), with
\(\mathbf{Q}_i'\mathbf{Q}_i = \mathbf{I}\). Note that
\(\mathbf{Q}_i'\mathbf{T}_i = \mathbf{0}\). From the observation that
\(\tilde{\mathbf{B}}_i\) can be written as \[
\tilde{\mathbf{B}}_i = \mathbf{I}_i - \mathbf{Q}_i \mathbf{Q}_i' + \mathbf{Q}_i \left(\mathbf{I} - \mathbf{R}_i \mathbf{M}_{\mathbf{\ddot{U}}} \mathbf{R}_i'\right)\mathbf{Q}_i',
\] it can be seen that \begin{equation}
\tilde{\mathbf{A}}_i = \tilde{\mathbf{B}}_i^{+1/2} = \mathbf{I}_i - \mathbf{Q}_i \mathbf{Q}_i' + \mathbf{Q}_i \left(\mathbf{I} - \mathbf{R}_i \mathbf{M}_{\mathbf{\ddot{U}}} \mathbf{R}_i'\right)^{+1/2} \mathbf{Q}_i'.
\end{equation} It follows that
\(\tilde{\mathbf{A}}_i \mathbf{T}_i = \mathbf{T}_i\).

Setting \begin{equation}
\mathbf{A}_i = \tilde{\mathbf{A}}_i - \mathbf{T}_i \mathbf{M_T}\mathbf{T}_i',
\end{equation} observe that \begin{align*}
\mathbf{B}_i \mathbf{A}_i \mathbf{B}_i \mathbf{A}_i &= \left(\tilde{\mathbf{B}}_i - \mathbf{T}_i \mathbf{M_T}\mathbf{T}_i'\right) \left(\tilde{\mathbf{A}}_i - \mathbf{T}_i \mathbf{M_T}\mathbf{T}_i'\right)\left(\tilde{\mathbf{B}}_i - \mathbf{T}_i \mathbf{M_T}\mathbf{T}_i'\right) \left(\tilde{\mathbf{A}}_i - \mathbf{T}_i \mathbf{M_T}\mathbf{T}_i'\right) \\
&= \left(\tilde{\mathbf{B}}_i\tilde{\mathbf{A}}_i - \mathbf{T}_i \mathbf{M_T}\mathbf{T}_i'\right)\left(\tilde{\mathbf{B}}_i\tilde{\mathbf{A}}_i - \mathbf{T}_i \mathbf{M_T}\mathbf{T}_i'\right) \\
&= \left(\tilde{\mathbf{B}}_i\tilde{\mathbf{A}}_i\tilde{\mathbf{B}}_i\tilde{\mathbf{A}}_i - \mathbf{T}_i \mathbf{M_T}\mathbf{T}_i'\right) \\
&= \left(\tilde{\mathbf{B}}_i - \mathbf{T}_i \mathbf{M_T}\mathbf{T}_i'\right) \\
&= \mathbf{B}_i.
\end{align*} It follows that \(\mathbf{A}_i\) is the symmetric square
root of the Moore-Penrose inverse of \(\mathbf{B}_i\), i.e.,
\(\mathbf{A}_i = \mathbf{B}_i^{+1/2}\). Because
\(\mathbf{T}_i ' \mathbf{\ddot{R}}_i= \mathbf{0}\), it can be seen that
\(\mathbf{A}_i \mathbf{\ddot{R}}_i = \left(\tilde{\mathbf{A}}_i - \mathbf{T}_i \mathbf{M_T}\mathbf{T}_i'\right)\mathbf{\ddot{R}}_i = \tilde{\mathbf{A}}_i \mathbf{\ddot{R}}_i\).
Finally, equality of \(\mathbf{\tilde{V}}^{CR}\) and \(\mathbf{V}^{CR}\)
follows by direct evaluation of Equation (\ref{eq:V_small}).

\newpage

\hypertarget{app:simulations}{%
\section{Details of simulation study}\label{app:simulations}}

We provide further details regarding the design of the simulations
reported in Section \ref{sec:simulation}. Table
\ref{tab:simulation_parameters} summarizes the factors manipulated in
the simulation.

\begin{table}[hbt]
\centering
\caption{Simulation design parameters} 
\label{tab:simulation_parameters}
\begin{tabular}{lc}
\toprule
Parameter & levels \\ \midrule
Design & RB, CR, DD \\
Balance & Balanced, Unbalanced \\
Outcome missingness & Complete data, 15\% missing \\
Clusters ($m$) & 15, 30, 50 \\
Units per cluster ($n$) & 12, 18, 30 \\
Intra-class correlation ($\tau^2$) & .05, .15, .25 \\
Treatment effect variability ($\sigma_\delta^2$) & .00, .04, .09 \\ 
Correlation among outcomes ($\rho$) & .2, .8 \\
\bottomrule
\end{tabular}
\end{table}

The simulations examined six distinct study designs. Outcomes are
measured for \(n\) units (which may be individuals, as in a
cluster-randomized or block-randomized design, or time-points, as in a
difference-in-differences panel) in each of \(m\) clusters under one of
three treatment conditions. Suppose that there are \(G\) sets of
clusters, each of size \(m_g\), where the clusters in each set have a
distinct configuration of treatment assignments. Let \(n_{ghi}\) denote
the number of units at which cluster \(i\) in configuration \(g\) is
observed under condition \(h\), for \(i=1,...,m\), \(g = 1,...,G\), and
\(h = 1,2,3\). Table \ref{tab:simulation_designs} summarizes the
cluster-level sample sizes and unit-level patterns of treatment
allocation for each of the six designs. The simulated designs included
the following:\\

\begin{enumerate}
\item A balanced, block-randomized design, with an un-equal allocation within each block. In the balanced design, the treatment allocation is identical for each block, so $G = 1$.
\item An unbalanced, block-randomized design, with two different patterns of treatment allocation ($G = 2$).
\item A balanced, cluster-randomized design, in which units are nested within clusters and an equal number of clusters are assigned to each treatment condition.
\item An unbalanced, cluster-randomized design, in which units are nested within clusters but the number of clusters assigned to each condition is not equal. 
\item A balanced difference-in-differences design with two patterns of treatment allocation ($G = 2$), in which half of the clusters are observed under the first treatment condition only and the remaining half are observed under all three conditions.
\item An unbalanced difference-in-differences design, again with two patterns of treatment allocation ($G = 2$), but where 2/3 of the clusters are observed under the first treatment condition only and the remaining $1 / 3$ of clusters are observed under all three conditions.
\end{enumerate}

\begin{landscape}

\begin{table}[H]
\centering
\caption{Study designs used for simulation} 
\label{tab:simulation_designs}
\begin{tabular}{llccc}
\toprule
Study design & Balance & Configuration & Clusters & Treatment allocation \\ 
\midrule
Randomized Block & Balanced & 1 & $m_1 = m$ & $n_{11i} = n / 2, n_{12i} = n / 3, n_{13i} = n / 6$ \\ \midrule
\multirow{2}{*}{Randomized Block} & \multirow{2}{*}{Unbalanced} & 1 & $m_1 = m / 2$ & $n_{11i} = n / 2, n_{12i} = n / 3, n_{13i} = n / 6$ \\
& & 2 & $m_2 = m / 2$ & $n_{21i} = n / 3, n_{22i} = 5n / 9, n_{23i} = n / 9$ \\ \midrule
\multirow{3}{*}{Cluster-Randomized} & \multirow{3}{*}{Balanced} & 1 & $m_1 = m / 3$ & $n_{11i} = n$ \\
& & 2 & $m_2 = m / 3$ & $n_{22i} = n$ \\ 
& & 3 & $m_3 = m / 3$ & $n_{33i} = n$ \\ \midrule
\multirow{3}{*}{Cluster-Randomized} & \multirow{3}{*}{Unbalanced} & 1 & $m_1 = m / 2$ & $n_{11i} = n$ \\
& & 2 & $m_2 = 3 m / 10$ & $n_{22i} = n$ \\ 
& & 3 & $m_3 = m / 5$ & $n_{33i} = n$ \\ \midrule
\multirow{2}{*}{Difference-in-Differences} & \multirow{2}{*}{Balanced} & 1 & $m_1 = m / 2$ & $n_{11i} = n$ \\
& & 2 & $m_2 = m / 2$ & $n_{21i} = n / 2, n_{22i} = n / 3, n_{23i} = n / 6$ \\ \midrule
\multirow{2}{*}{Difference-in-Differences} & \multirow{2}{*}{Unbalanced} & 1 & $m_1 = 2m / 3$ & $n_{11i} = n$ \\
& & 2 & $m_2 = m / 3$ & $n_{21i} = n / 2, n_{22i} = n / 3, n_{23i} = n / 6$ \\ 
\bottomrule
\end{tabular}
\end{table}

\newpage

\section{Additional simulation results}
\label{app:sim-results}

\subsection{Rejection rates of AHT and standard tests}

\begin{figure}[H]

{\centering \includegraphics[width=\linewidth]{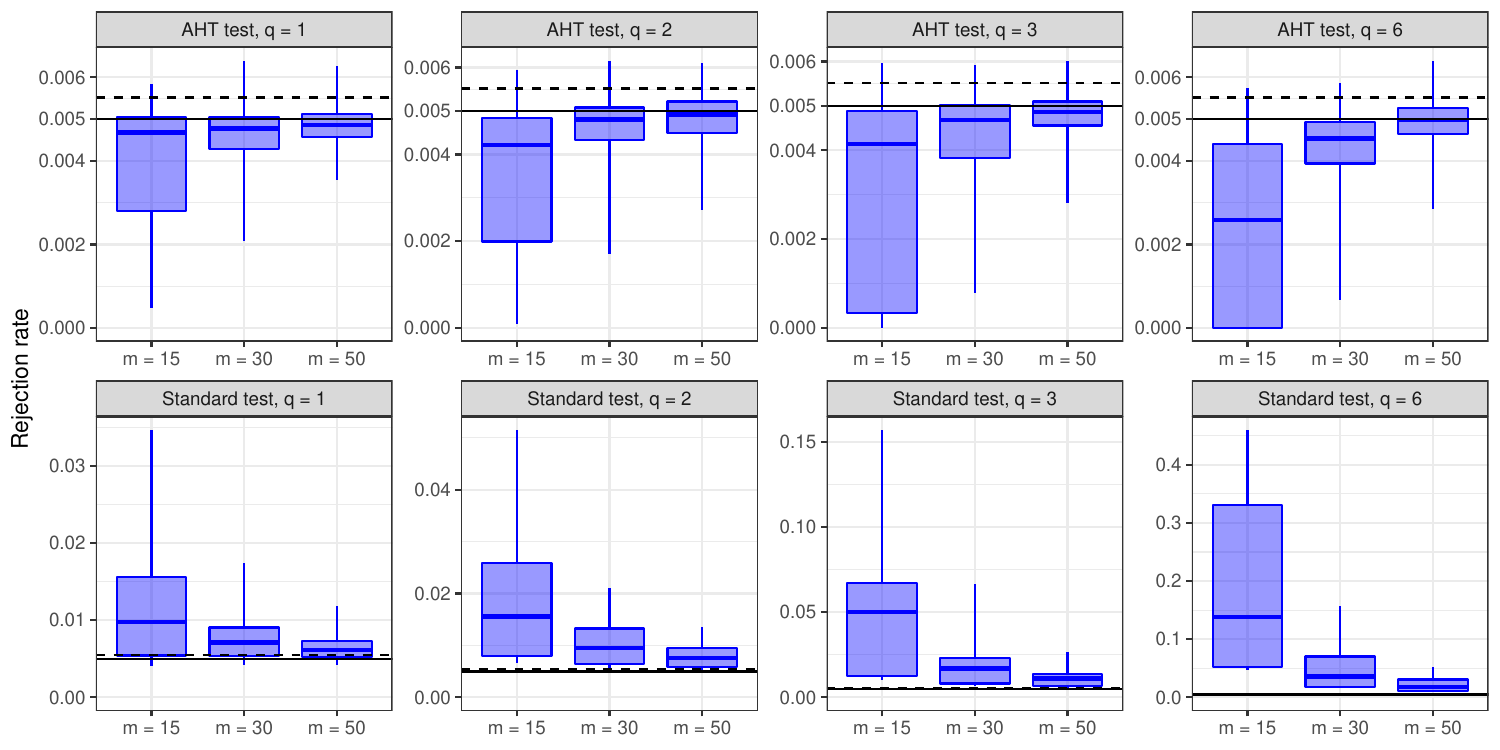} 

}

\caption{Rejection rates of AHT and standard tests for $\alpha = .005$, by dimension of hypothesis ($q$) and sample size ($m$).}\label{fig:overview_005}
\end{figure}

\begin{figure}[H]

{\centering \includegraphics[width=\linewidth]{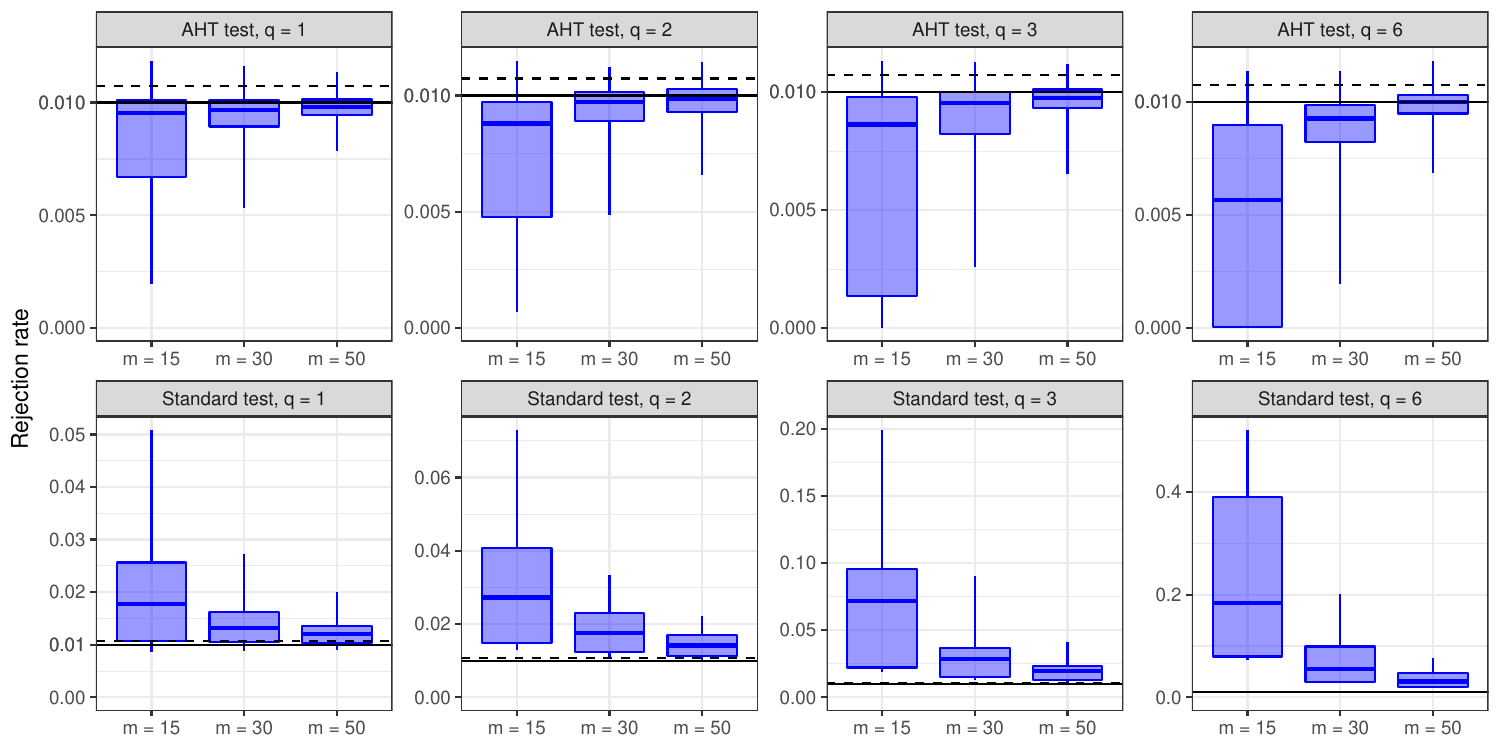} 

}

\caption{Rejection rates of AHT and standard tests for $\alpha = .01$, by dimension of hypothesis ($q$) and sample size ($m$).}\label{fig:overview_01}
\end{figure}

\begin{figure}[H]

{\centering \includegraphics[width=\linewidth]{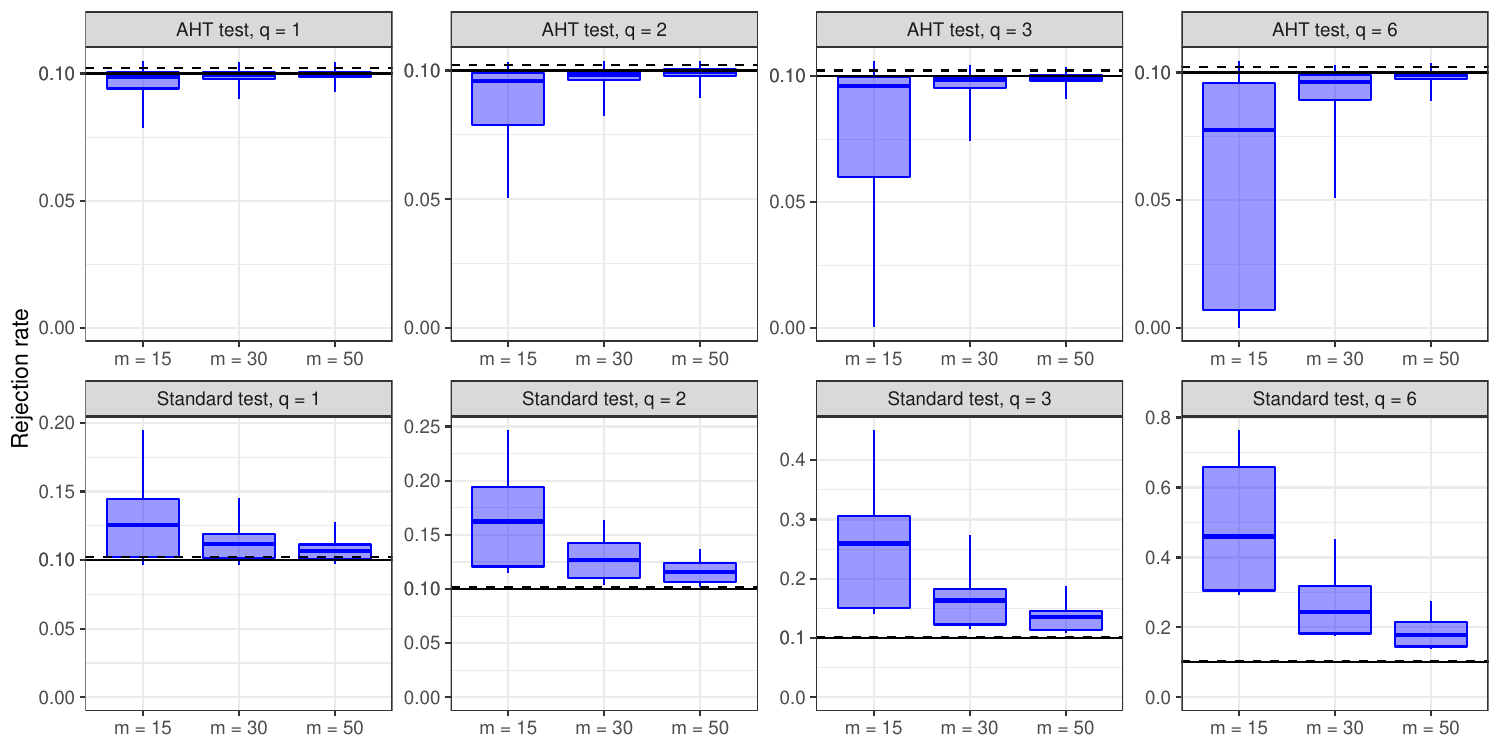} 

}

\caption{Rejection rates of AHT and standard tests for $\alpha = .10$, by dimension of hypothesis ($q$) and sample size ($m$).}\label{fig:overview_10}
\end{figure}

\subsection{Rejection rates of AHT and standard tests by study design}

\begin{figure}[H]

{\centering \includegraphics[width=\linewidth]{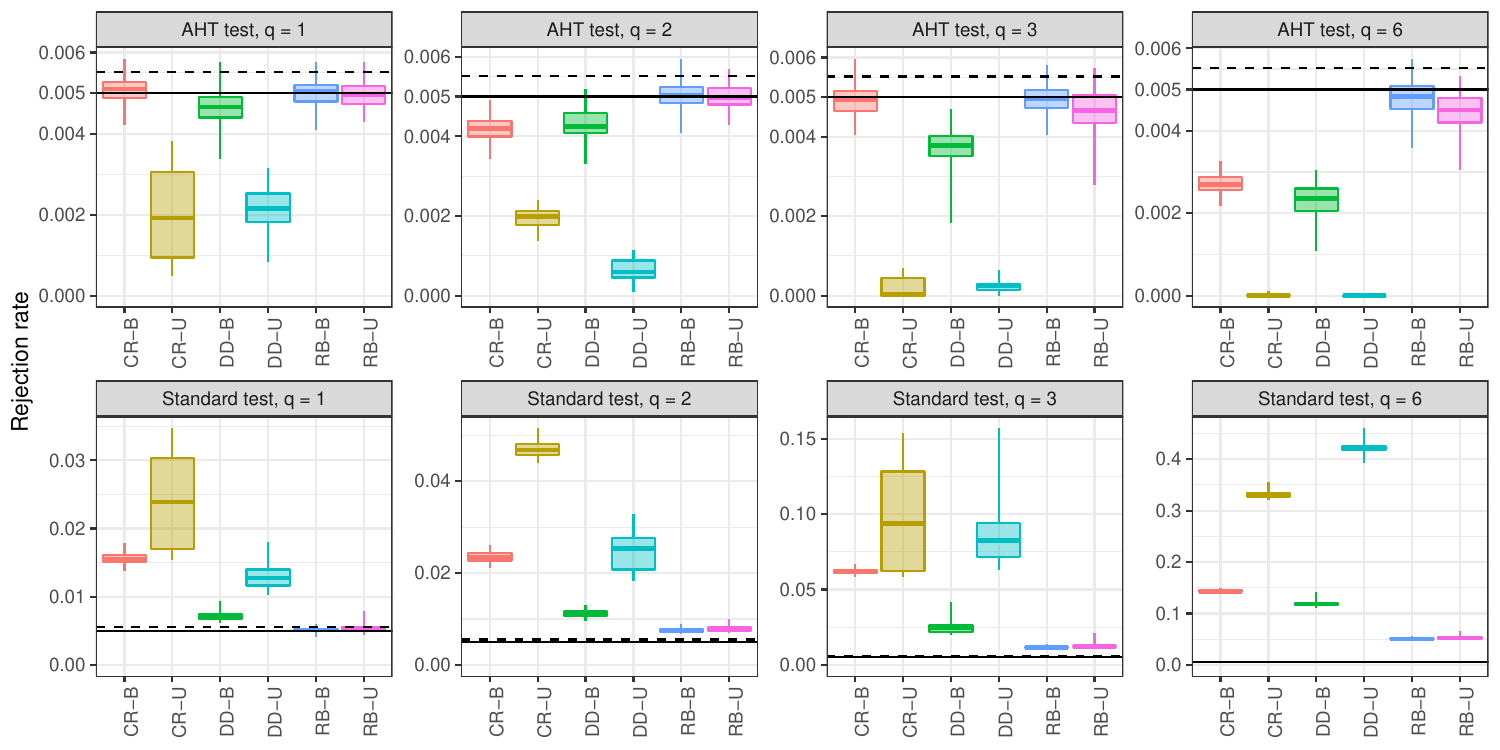} 

}

\caption{Rejection rates of AHT and standard tests, by study design and dimension of hypothesis ($q$) for $\alpha = .005$ and $m = 15$. CR = cluster-randomized design; DD = difference-in-differences design; RB = randomized block design; B = balanced; U = unbalanced.}\label{fig:balance_005_15}
\end{figure}

\begin{figure}[H]

{\centering \includegraphics[width=\linewidth]{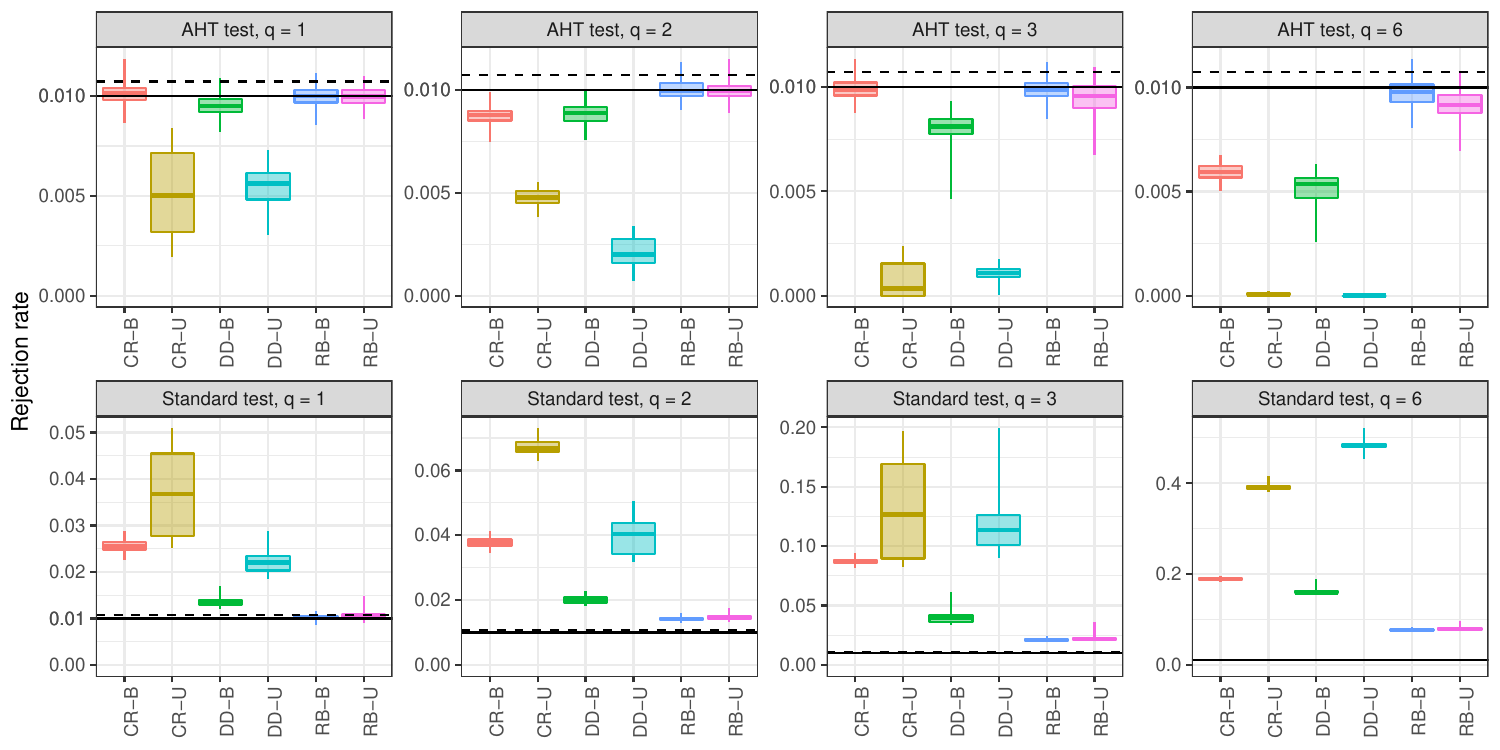} 

}

\caption{Rejection rates of AHT and standard tests, by study design and dimension of hypothesis ($q$) for $\alpha = .01$ and $m = 15$. CR = cluster-randomized design; DD = difference-in-differences design; RB = randomized block design; B = balanced; U = unbalanced.}\label{fig:balance_01_15}
\end{figure}

\begin{figure}[H]

{\centering \includegraphics[width=\linewidth]{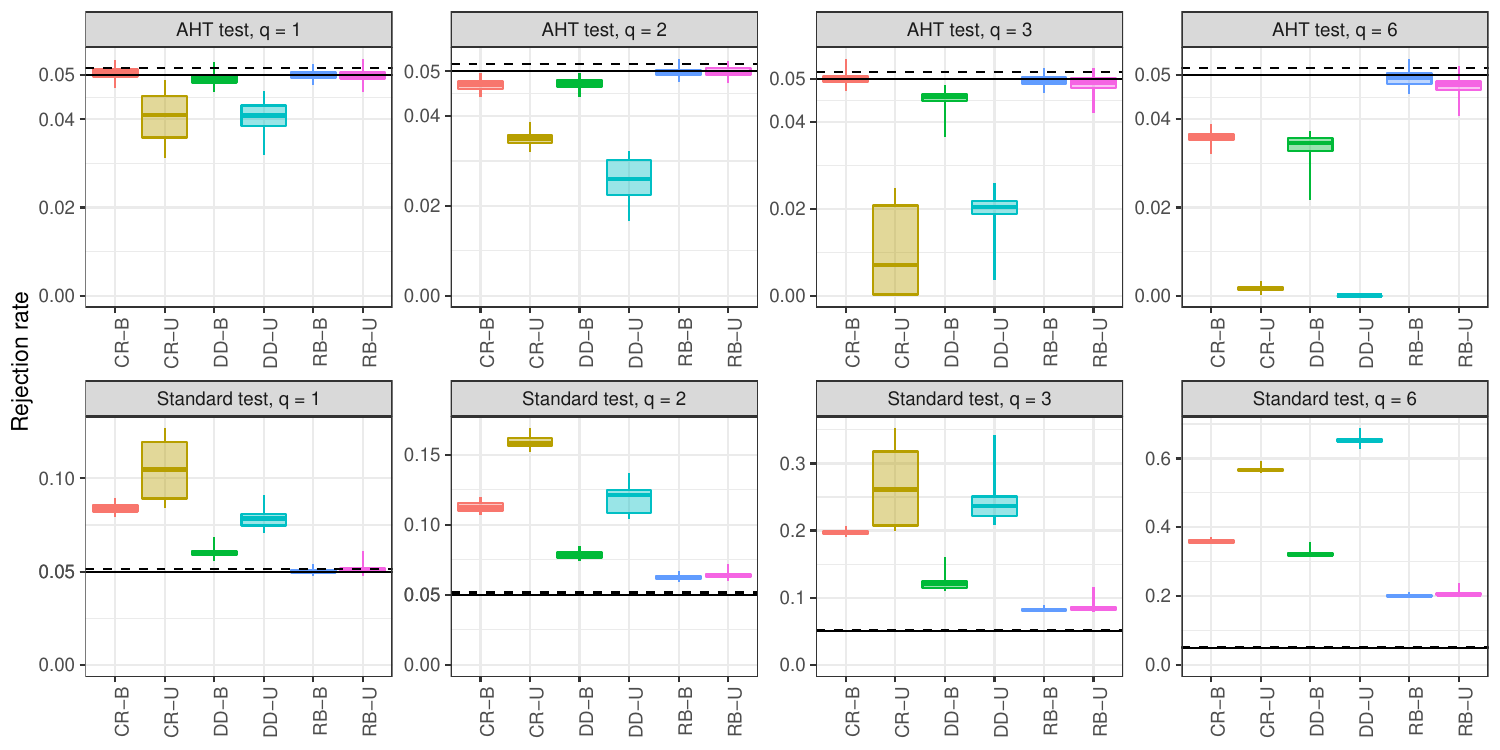} 

}

\caption{Rejection rates of AHT and standard tests, by study design and dimension of hypothesis ($q$) for $\alpha = .05$ and $m = 15$. CR = cluster-randomized design; DD = difference-in-differences design; RB = randomized block design; B = balanced; U = unbalanced.}\label{fig:balance_05_15}
\end{figure}

\begin{figure}[H]

{\centering \includegraphics[width=\linewidth]{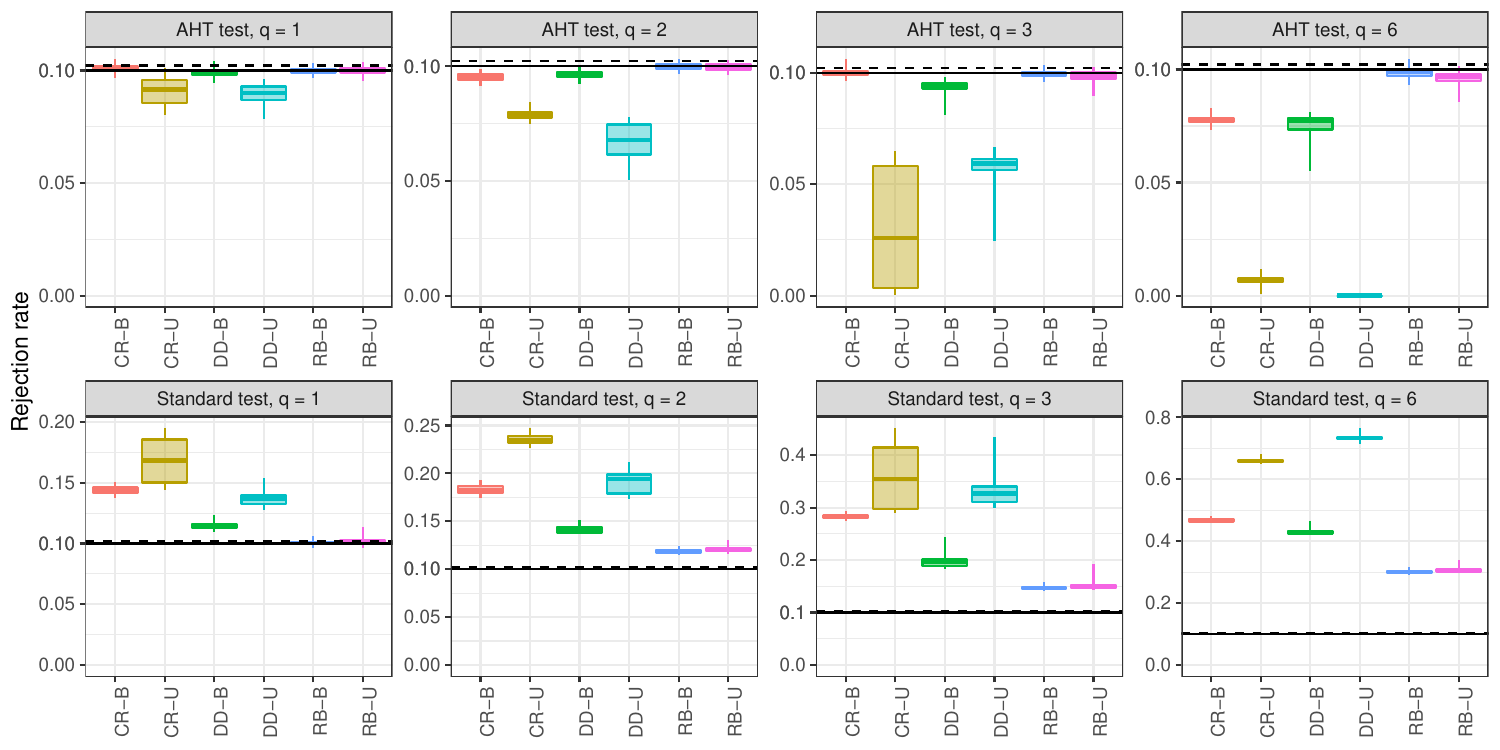} 

}

\caption{Rejection rates of AHT and standard tests, by study design and dimension of hypothesis ($q$) for $\alpha = .10$ and $m = 15$. CR = cluster-randomized design; DD = difference-in-differences design; RB = randomized block design; B = balanced; U = unbalanced.}\label{fig:balance_10_15}
\end{figure}

\begin{figure}[H]

{\centering \includegraphics[width=\linewidth]{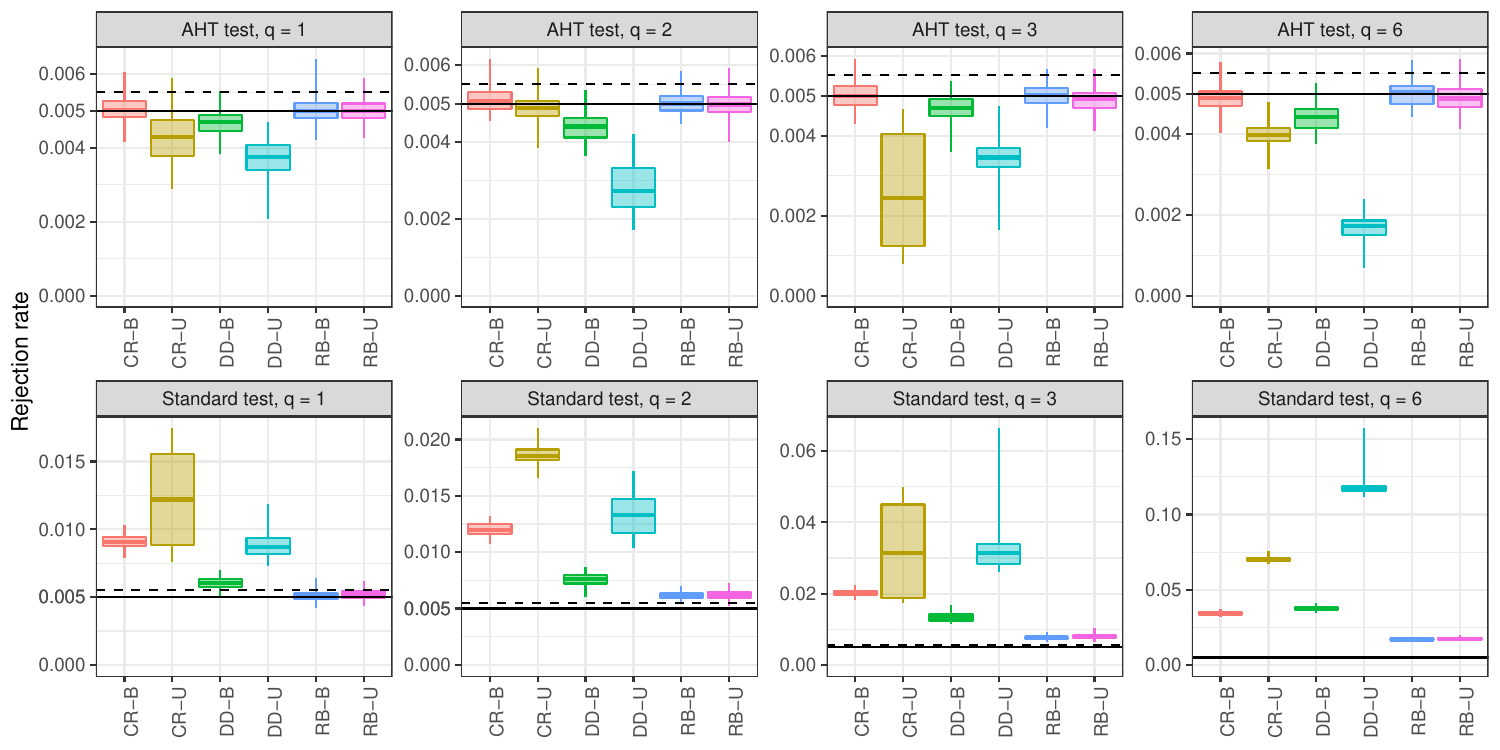} 

}

\caption{Rejection rates of AHT and standard tests, by study design and dimension of hypothesis ($q$) for $\alpha = .005$ and $m = 30$. CR = cluster-randomized design; DD = difference-in-differences design; RB = randomized block design; B = balanced; U = unbalanced.}\label{fig:balance_005_30}
\end{figure}

\begin{figure}[H]

{\centering \includegraphics[width=\linewidth]{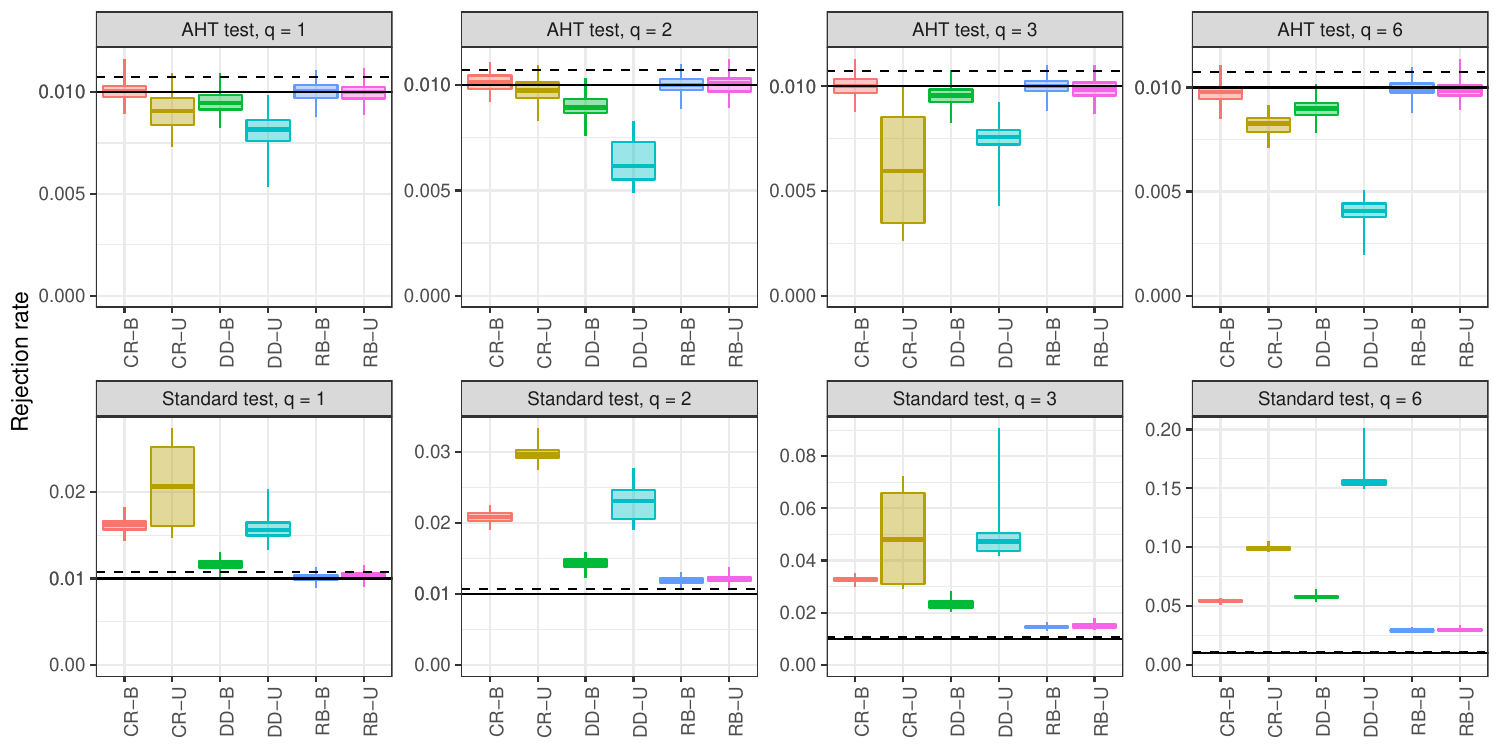} 

}

\caption{Rejection rates of AHT and standard tests, by study design and dimension of hypothesis ($q$) for $\alpha = .01$ and $m = 30$. CR = cluster-randomized design; DD = difference-in-differences design; RB = randomized block design; B = balanced; U = unbalanced.}\label{fig:balance_01_30}
\end{figure}

\begin{figure}[H]

{\centering \includegraphics[width=\linewidth]{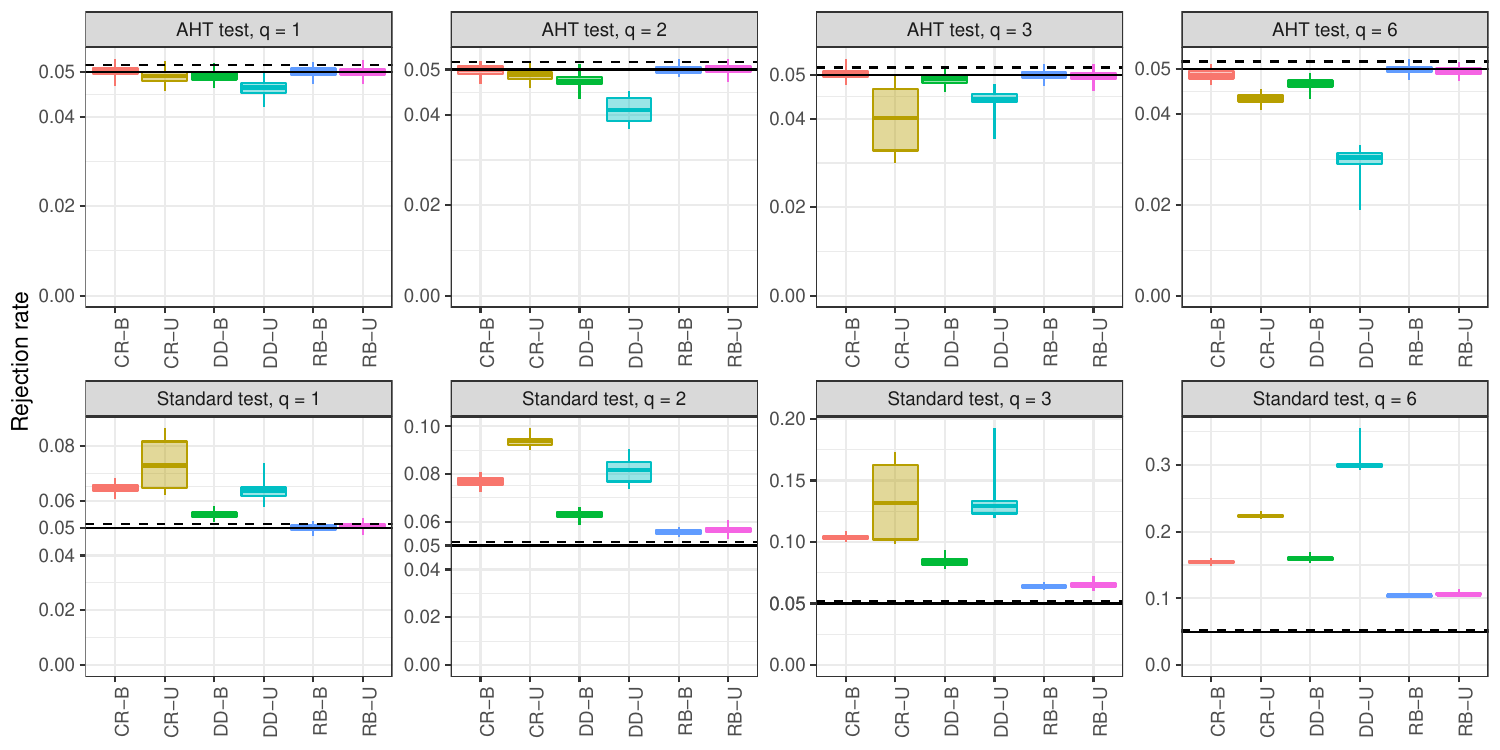} 

}

\caption{Rejection rates of AHT and standard tests, by study design and dimension of hypothesis ($q$) for $\alpha = .05$ and $m = 30$. CR = cluster-randomized design; DD = difference-in-differences design; RB = randomized block design; B = balanced; U = unbalanced.}\label{fig:balance_05_30}
\end{figure}

\begin{figure}[H]

{\centering \includegraphics[width=\linewidth]{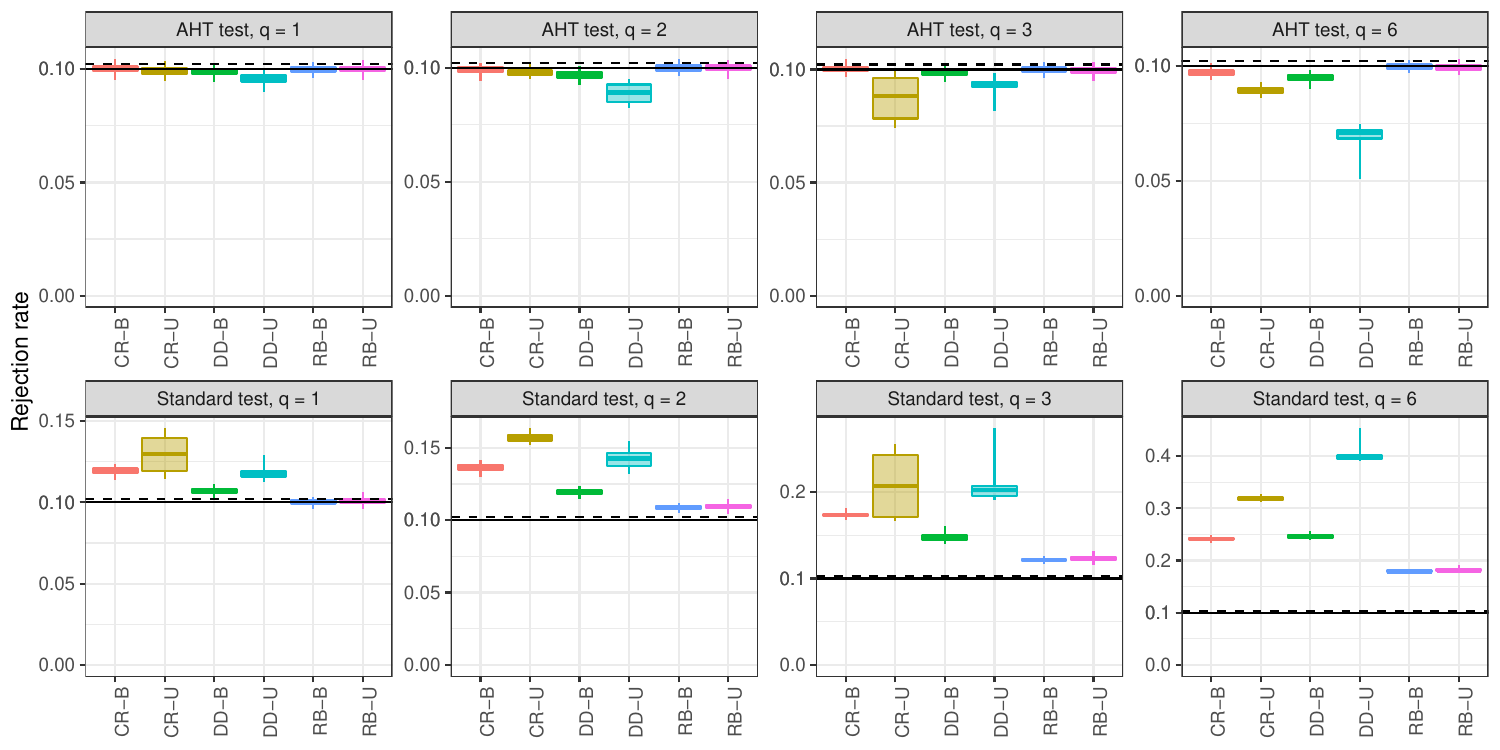} 

}

\caption{Rejection rates of AHT and standard tests, by study design and dimension of hypothesis ($q$) for $\alpha = .10$ and $m = 30$. CR = cluster-randomized design; DD = difference-in-differences design; RB = randomized block design; B = balanced; U = unbalanced.}\label{fig:balance_10_30}
\end{figure}

\begin{figure}[H]

{\centering \includegraphics[width=\linewidth]{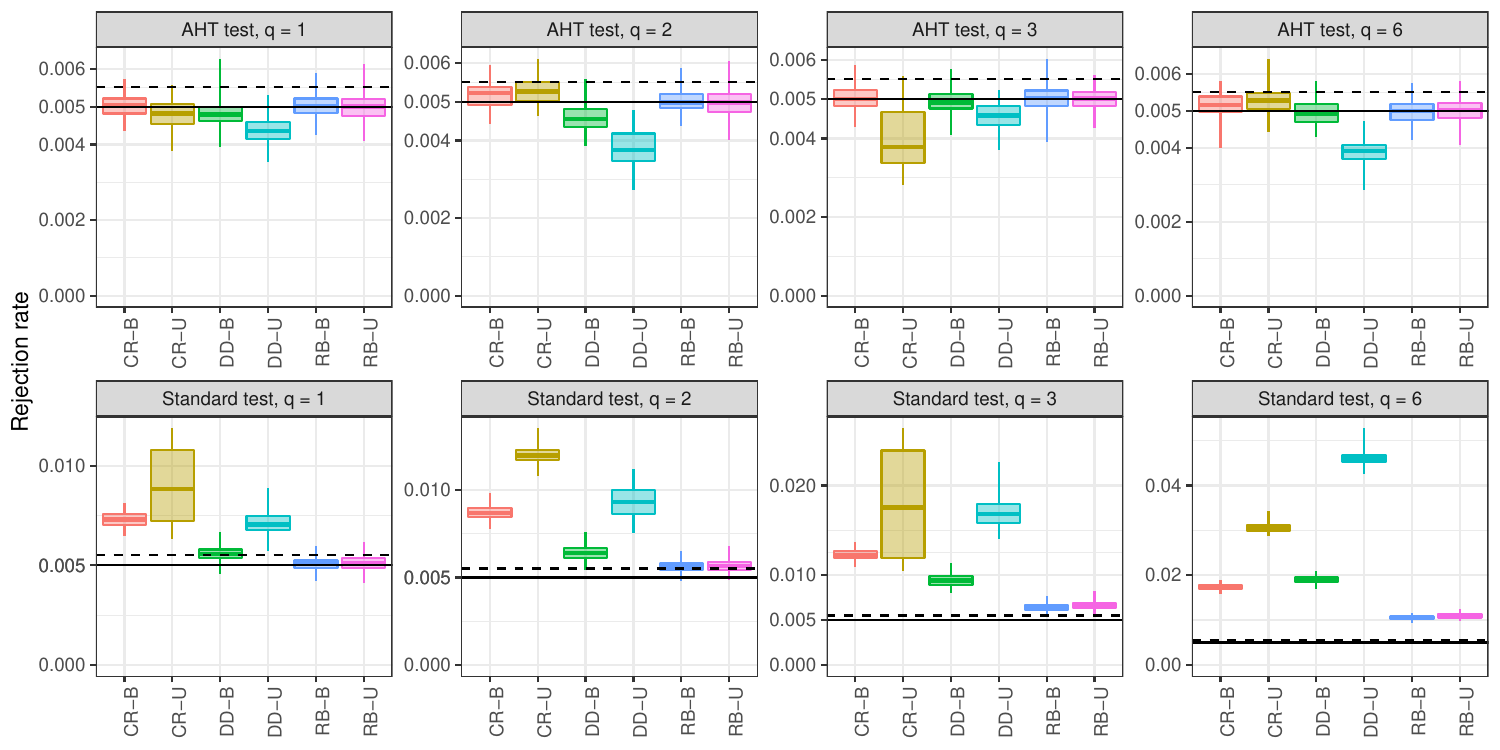} 

}

\caption{Rejection rates of AHT and standard tests, by study design and dimension of hypothesis ($q$) for $\alpha = .005$ and $m = 50$. CR = cluster-randomized design; DD = difference-in-differences design; RB = randomized block design; B = balanced; U = unbalanced.}\label{fig:balance_005_50}
\end{figure}

\begin{figure}[H]

{\centering \includegraphics[width=\linewidth]{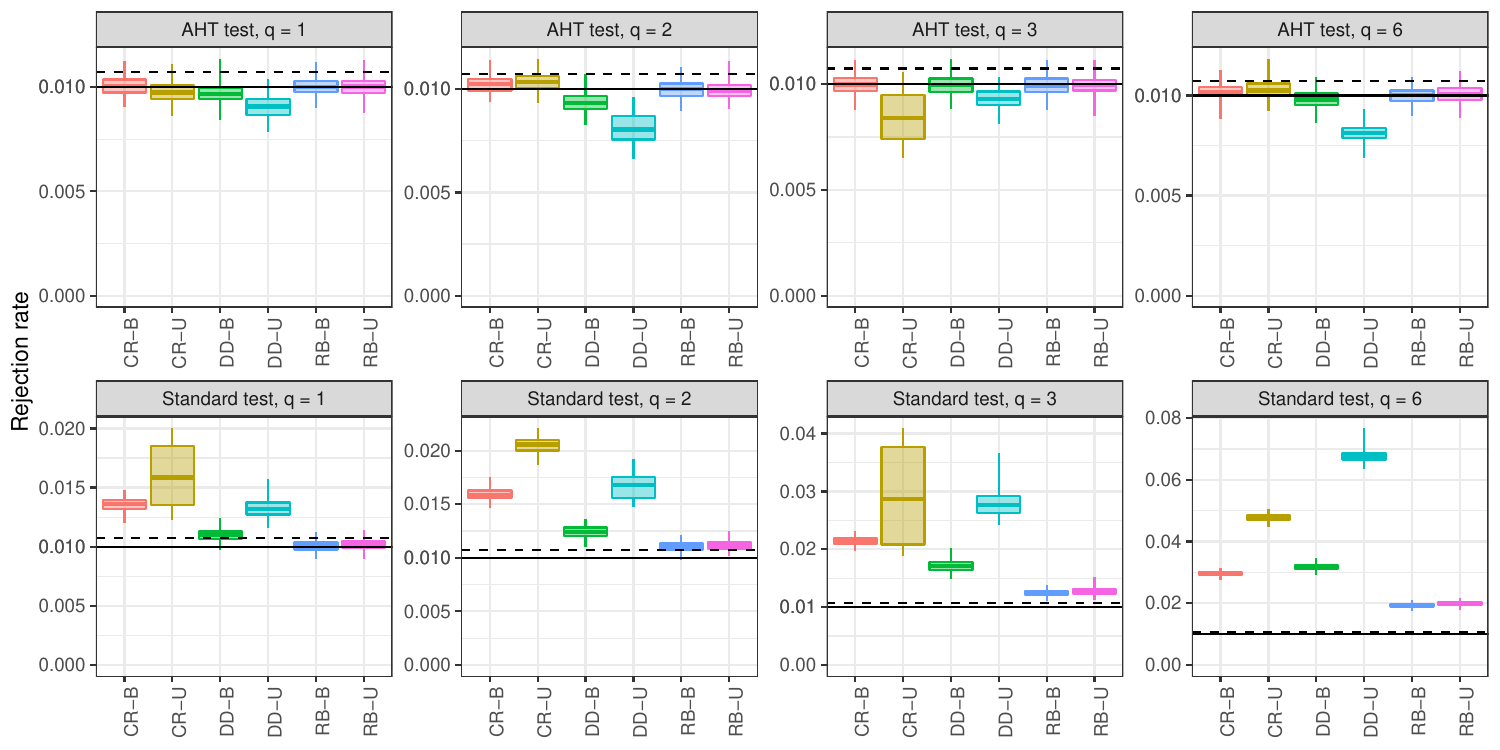} 

}

\caption{Rejection rates of AHT and standard tests, by study design and dimension of hypothesis ($q$) for $\alpha = .01$ and $m = 50$. CR = cluster-randomized design; DD = difference-in-differences design; RB = randomized block design; B = balanced; U = unbalanced.}\label{fig:balance_01_50}
\end{figure}

\begin{figure}[H]

{\centering \includegraphics[width=\linewidth]{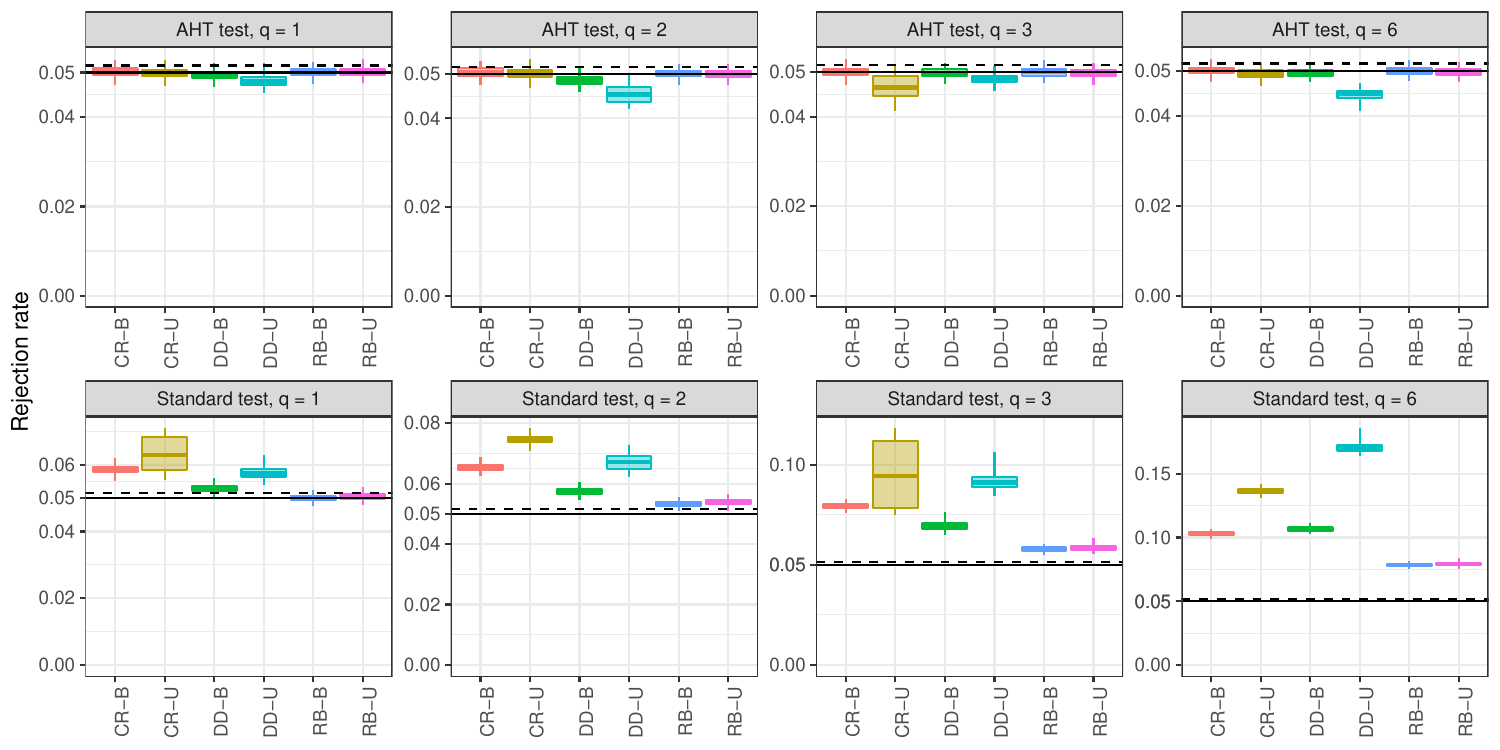} 

}

\caption{Rejection rates of AHT and standard tests, by study design and dimension of hypothesis ($q$) for $\alpha = .05$ and $m = 50$. CR = cluster-randomized design; DD = difference-in-differences design; RB = randomized block design; B = balanced; U = unbalanced.}\label{fig:balance_05_50}
\end{figure}

\begin{figure}[H]

{\centering \includegraphics[width=\linewidth]{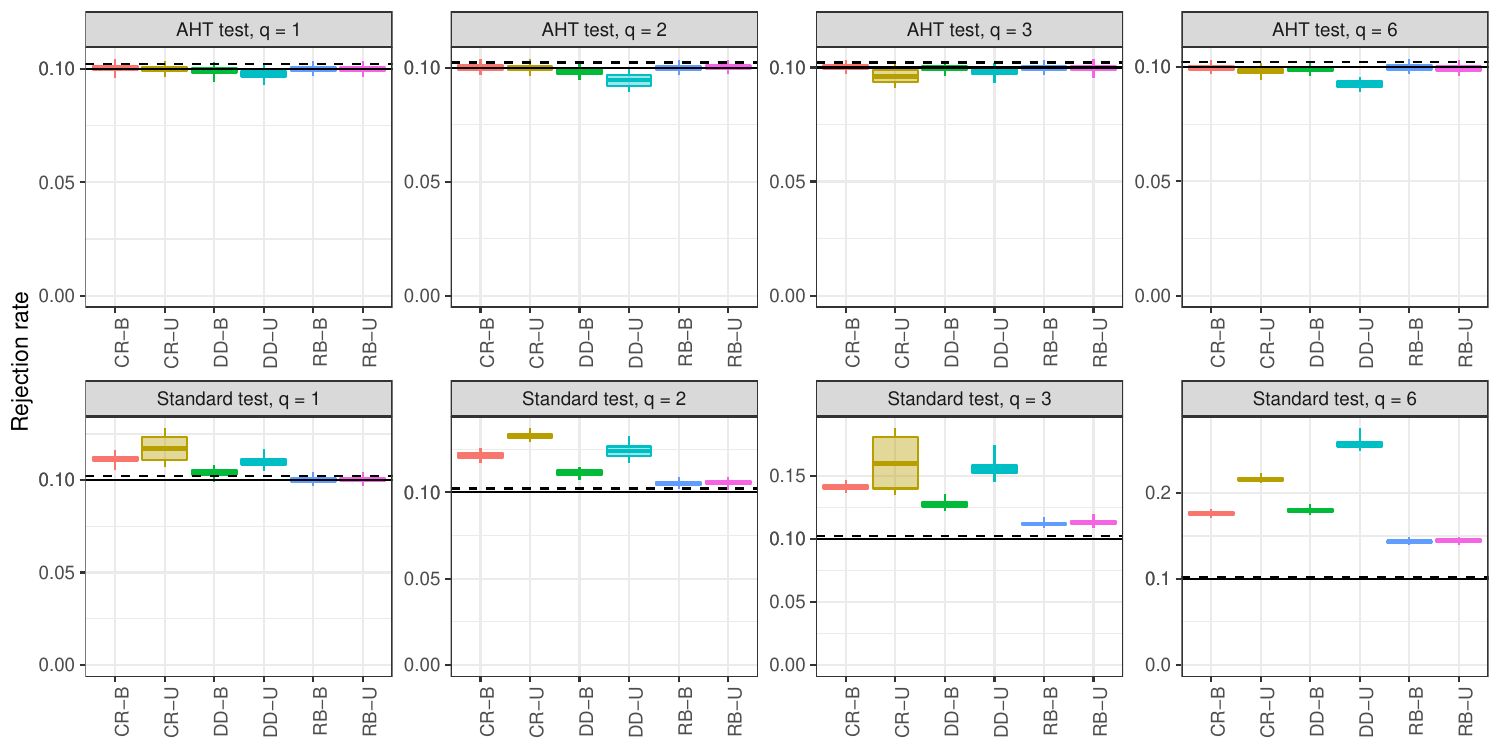} 

}

\caption{Rejection rates of AHT and standard tests, by study design and dimension of hypothesis ($q$) for $\alpha = .10$ and $m = 50$. CR = cluster-randomized design; DD = difference-in-differences design; RB = randomized block design; B = balanced; U = unbalanced.}\label{fig:balance_10_50}
\end{figure}

\subsection{Rejection rates of AHT test using CR1 or CR2, with and without accounting for absorption}

\begin{figure}[H]

{\centering \includegraphics[width=\linewidth]{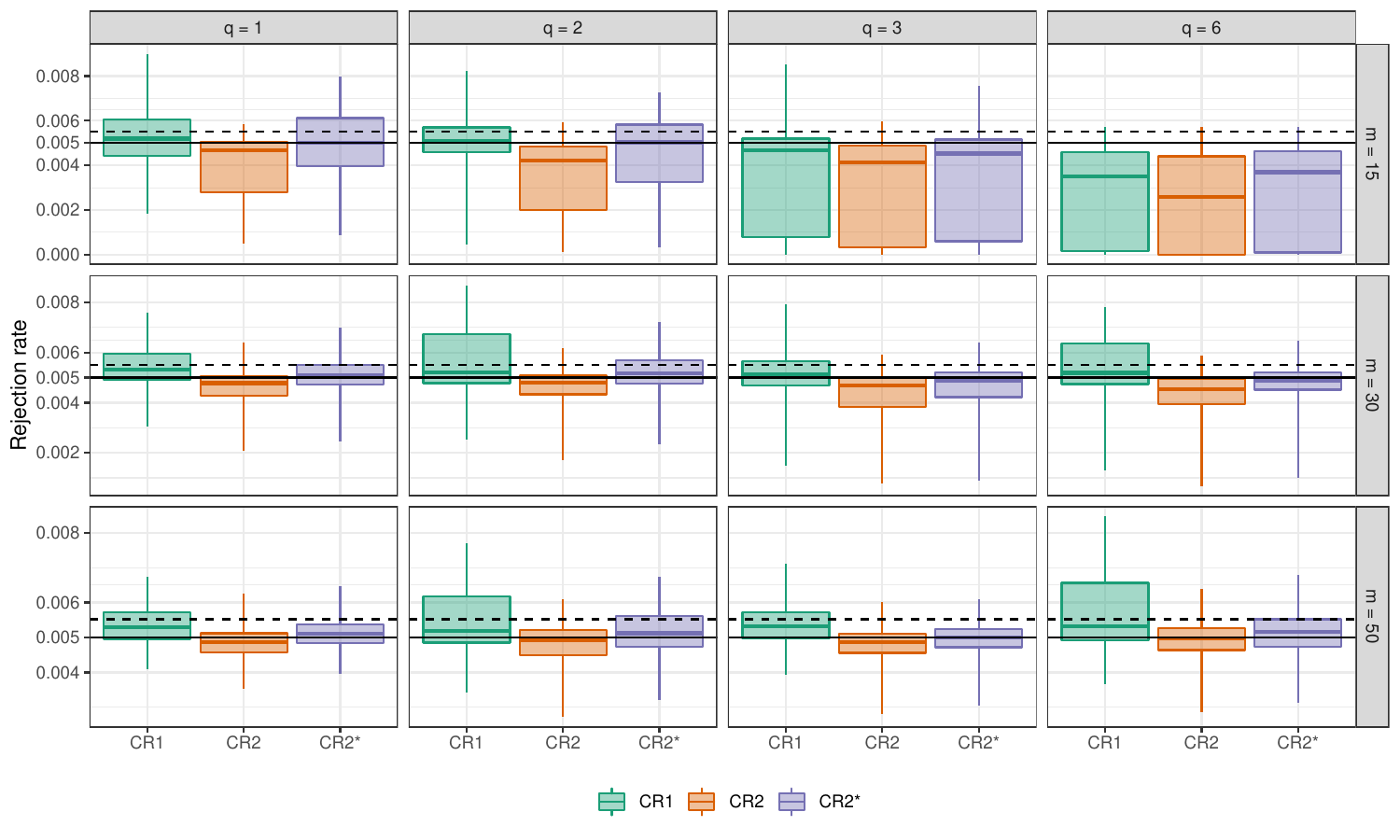} 

}

\caption{Rejection rates of AHT tests using CR1, CR2, or CR2 calculated without accounting for absorption of fixed effects (CR2*), by sample size ($m$) and dimension of hypothesis ($q$), for $\alpha = .005$.}\label{fig:absorption_005}
\end{figure}

\begin{figure}[H]

{\centering \includegraphics[width=\linewidth]{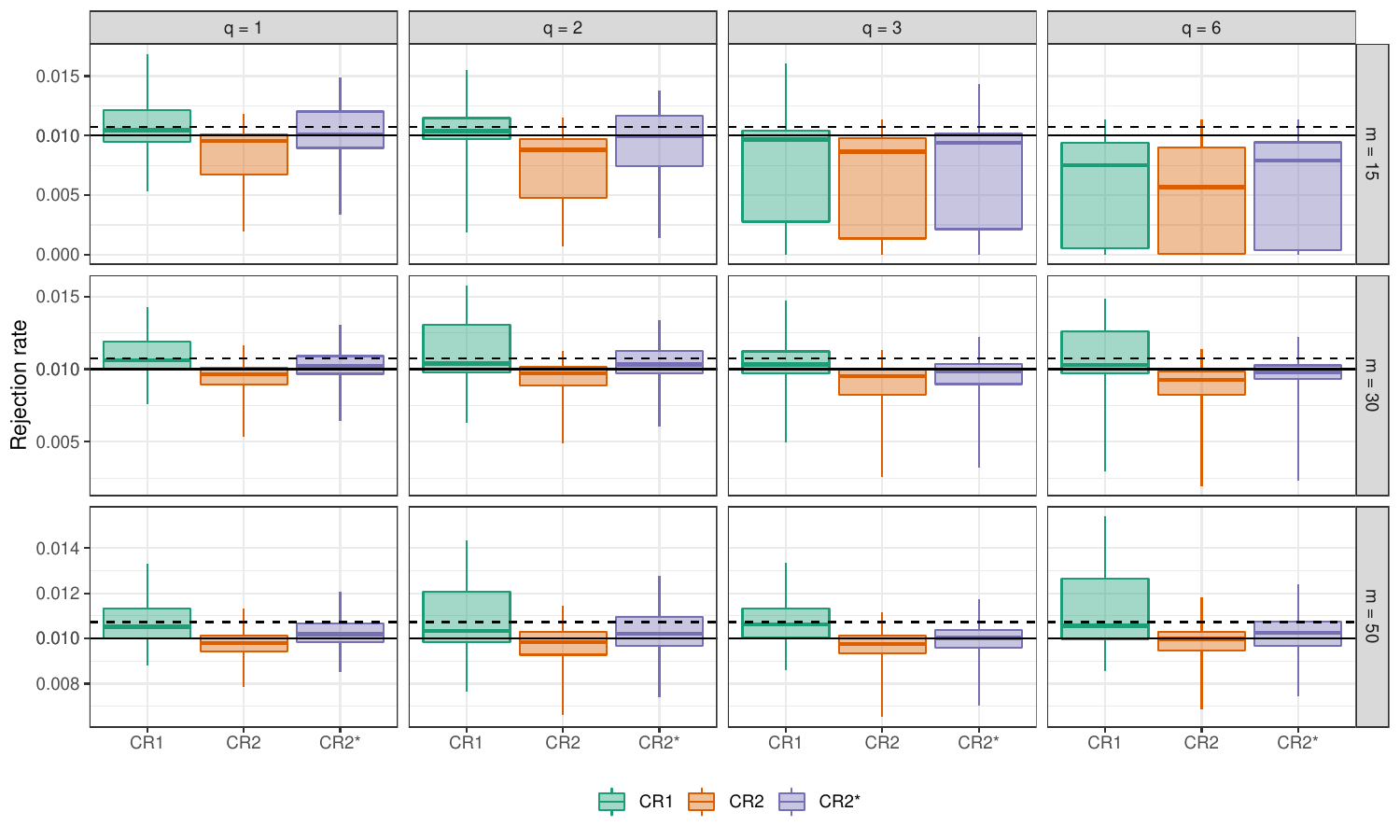} 

}

\caption{Rejection rates of AHT tests using CR1, CR2, or CR2 calculated without accounting for absorption of fixed effects (CR2*), by sample size ($m$) and dimension of hypothesis ($q$), for $\alpha = .01$.}\label{fig:absorption_01}
\end{figure}

\begin{figure}[H]

{\centering \includegraphics[width=\linewidth]{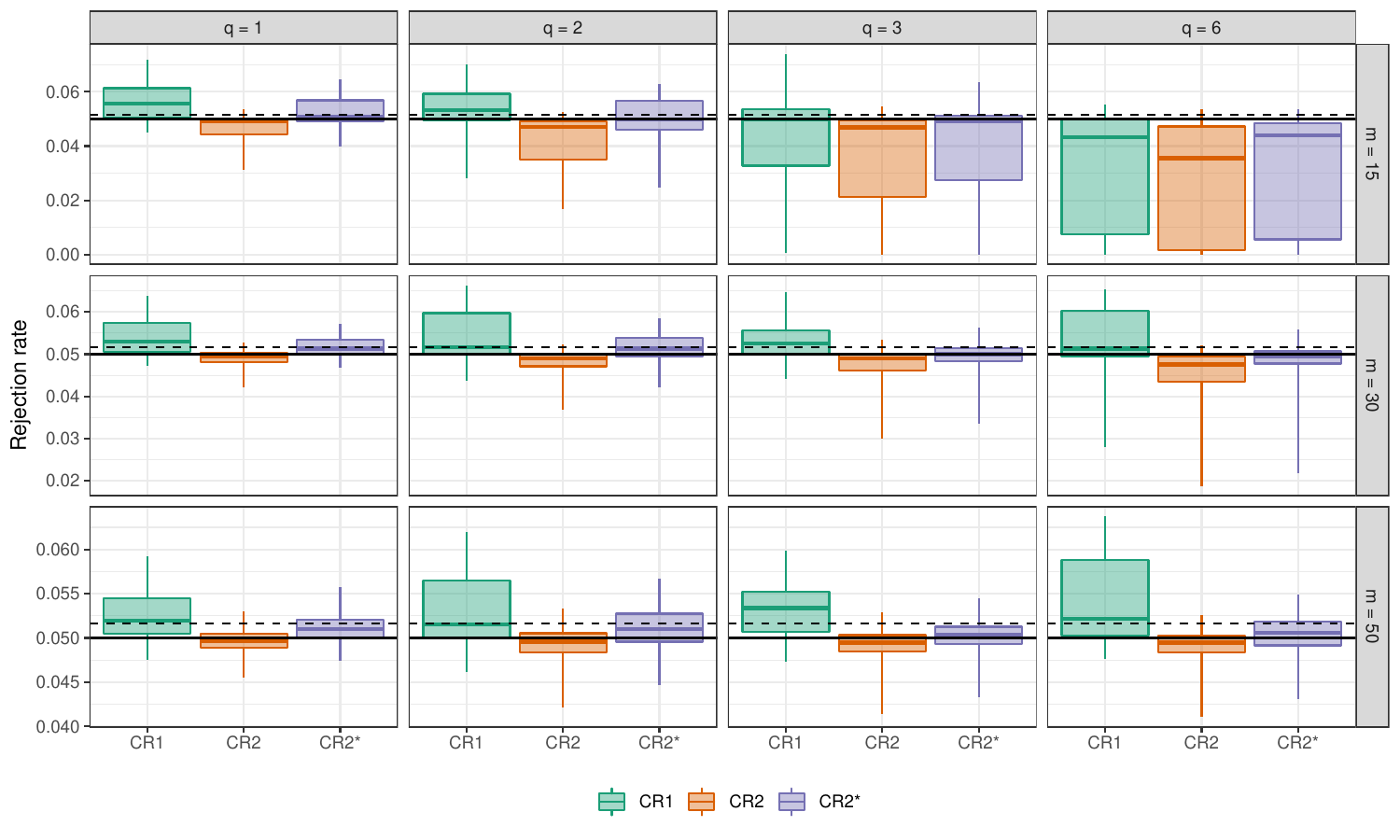} 

}

\caption{Rejection rates of AHT tests using CR1, CR2, or CR2 calculated without accounting for absorption of fixed effects (CR2*), by sample size ($m$) and dimension of hypothesis ($q$), for $\alpha = .05$.}\label{fig:absorption_05}
\end{figure}

\begin{figure}[H]

{\centering \includegraphics[width=\linewidth]{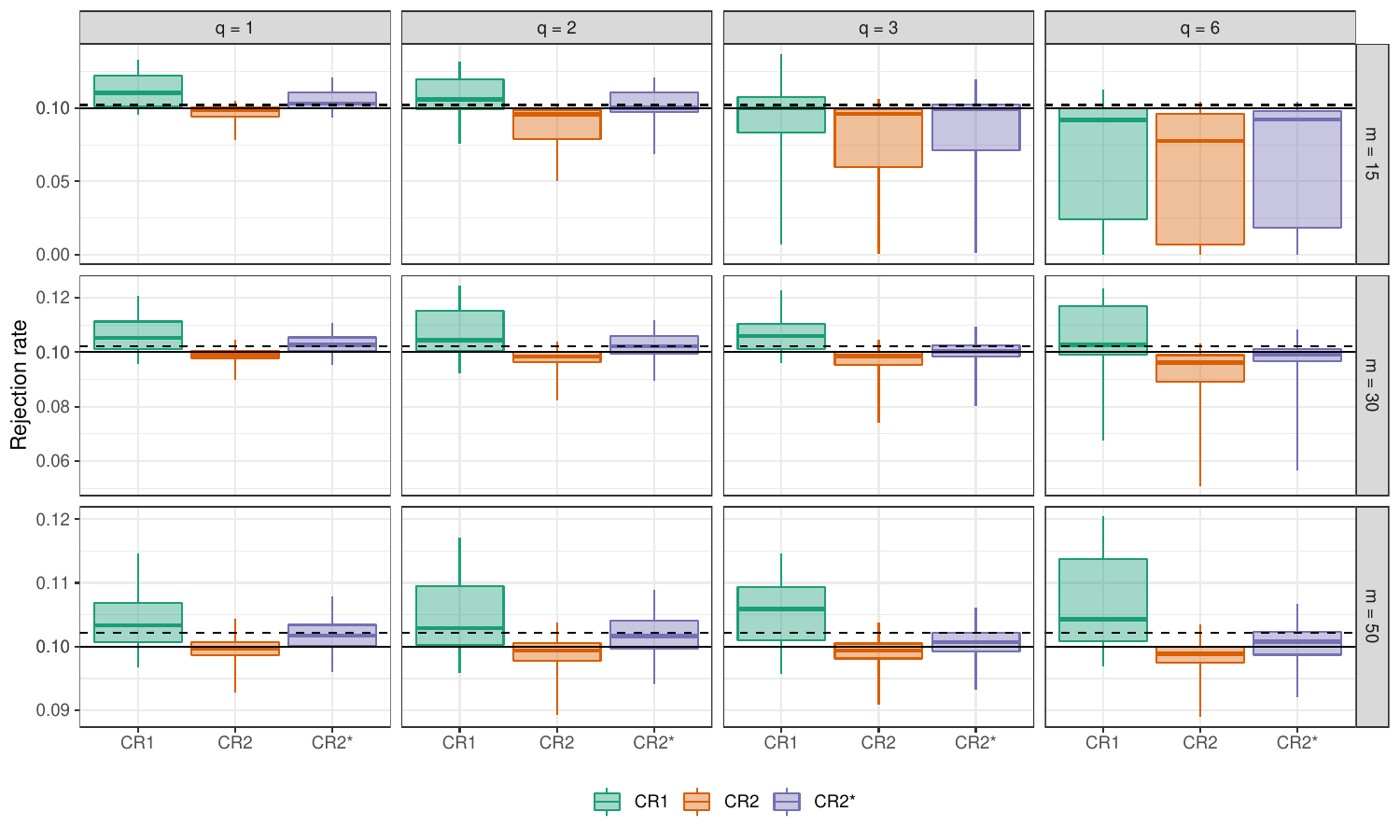} 

}

\caption{Rejection rates of AHT tests using CR1, CR2, or CR2 calculated without accounting for absorption of fixed effects (CR2*), by sample size ($m$) and dimension of hypothesis ($q$), for $\alpha = .10$.}\label{fig:absorption_10}
\end{figure}

\subsection{Rejection rates of AHT test by degree of working model misspecification}

\begin{figure}[H]

{\centering \includegraphics[width=\linewidth]{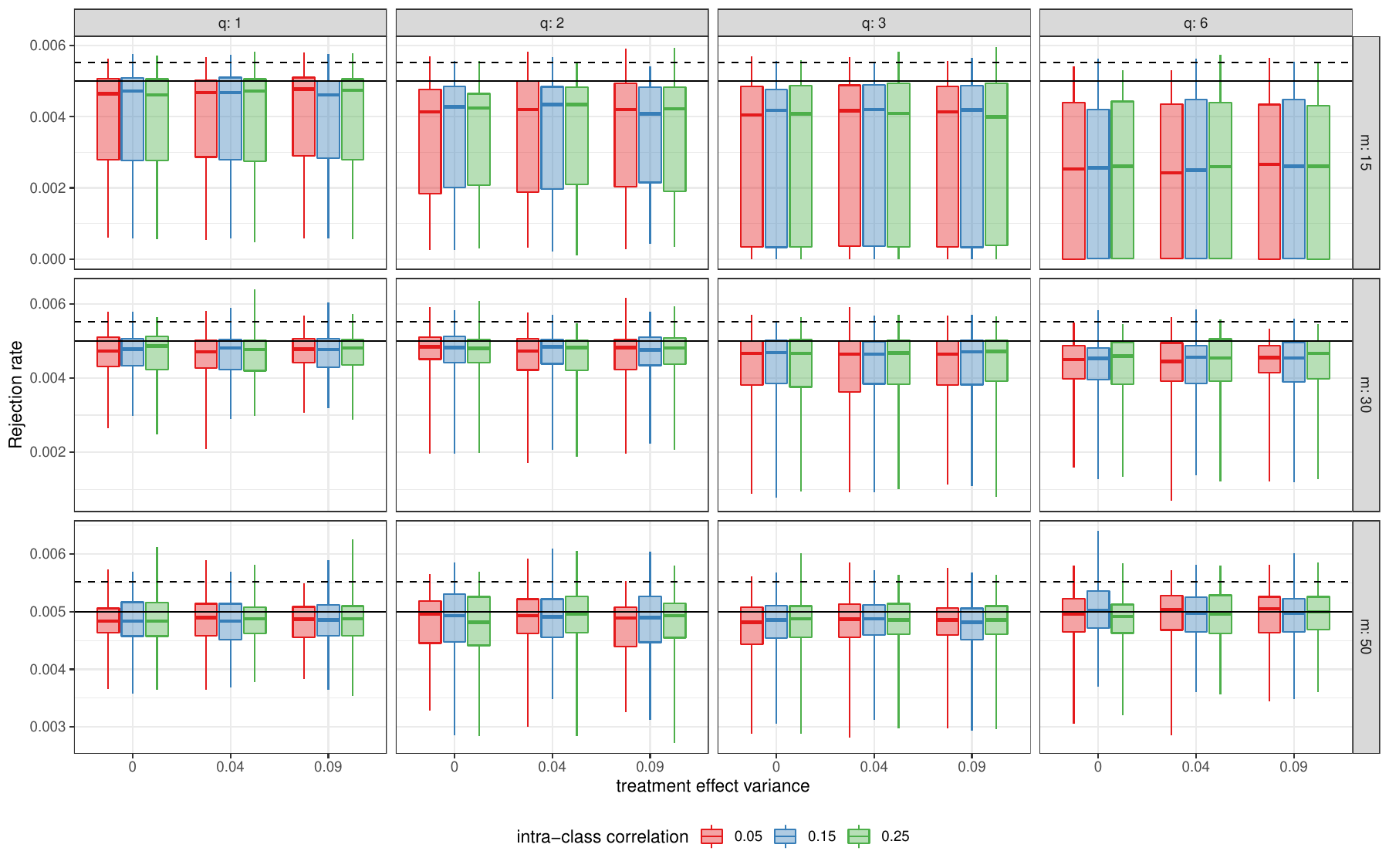} 

}

\caption{Rejection rates of CR2 AHT test, by treatment effect variance and intra-class correlation for $\alpha = .005$.}\label{fig:misspecification_005}
\end{figure}

\begin{figure}[H]

{\centering \includegraphics[width=\linewidth]{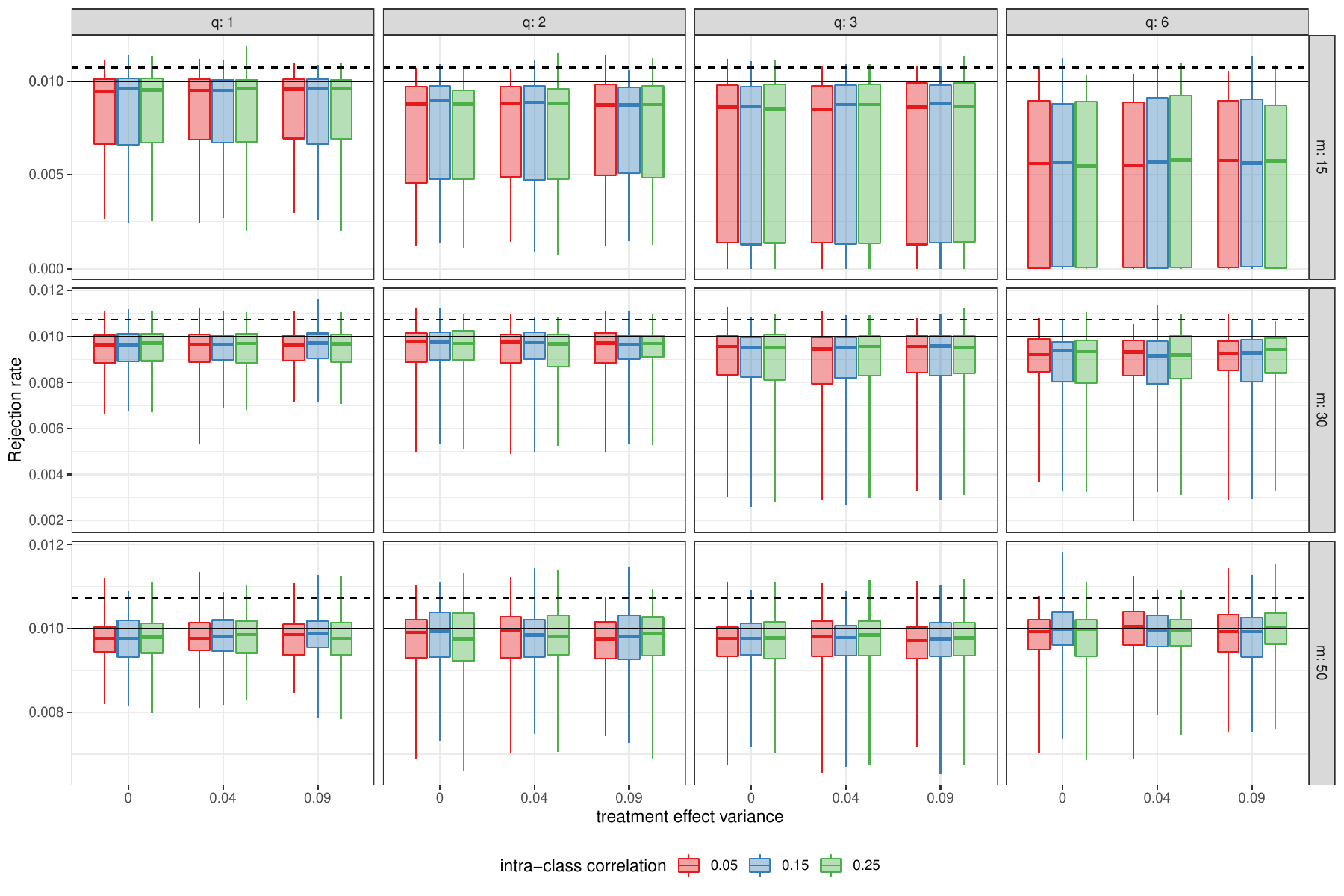} 

}

\caption{Rejection rates of CR2 AHT test, by treatment effect variance and intra-class correlation for $\alpha = .01$.}\label{fig:misspecification_01}
\end{figure}

\begin{figure}[H]

{\centering \includegraphics[width=\linewidth]{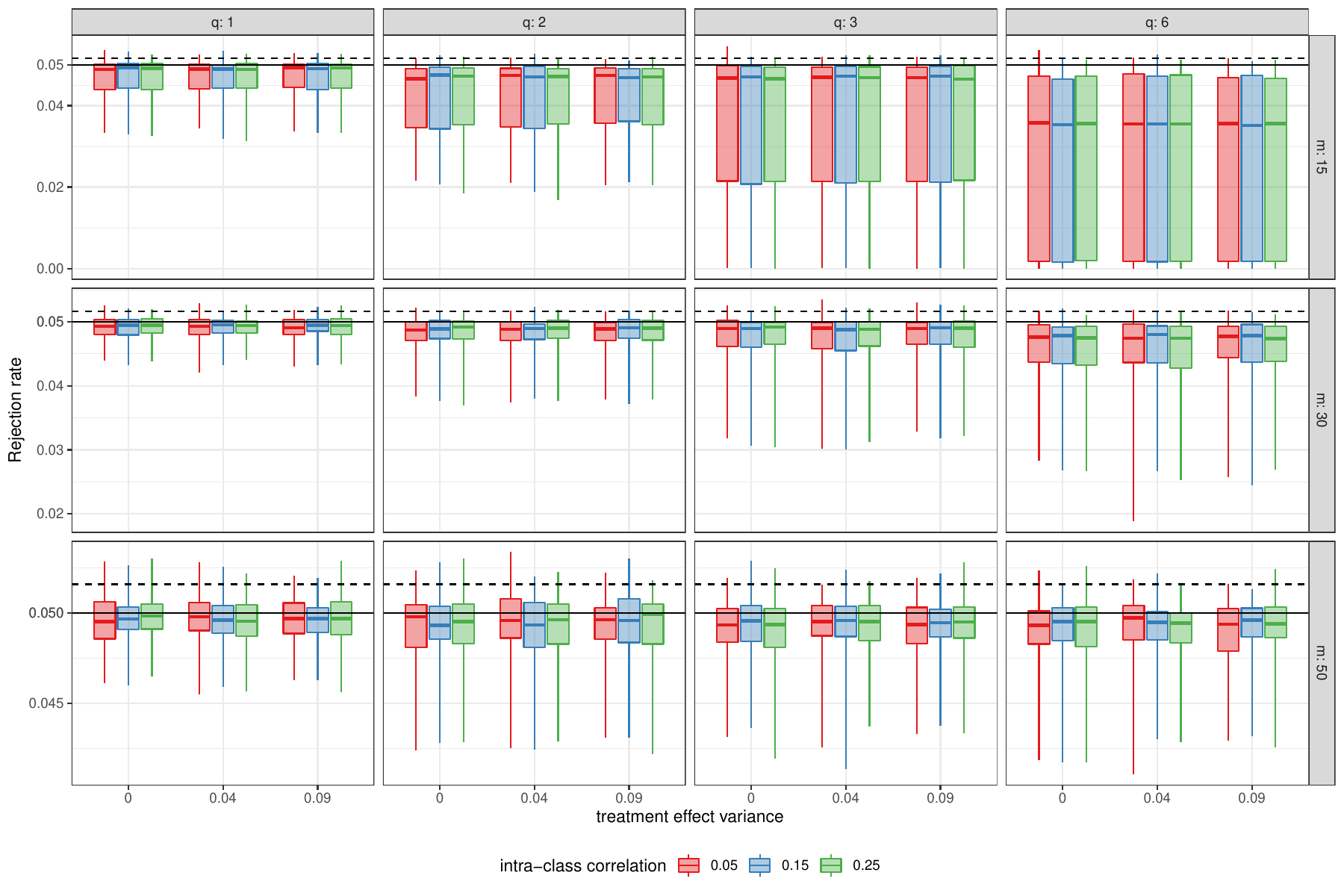} 

}

\caption{Rejection rates of CR2 AHT test, by treatment effect variance and intra-class correlation for $\alpha = .05$.}\label{fig:misspecification_05}
\end{figure}

\begin{figure}[H]

{\centering \includegraphics[width=\linewidth]{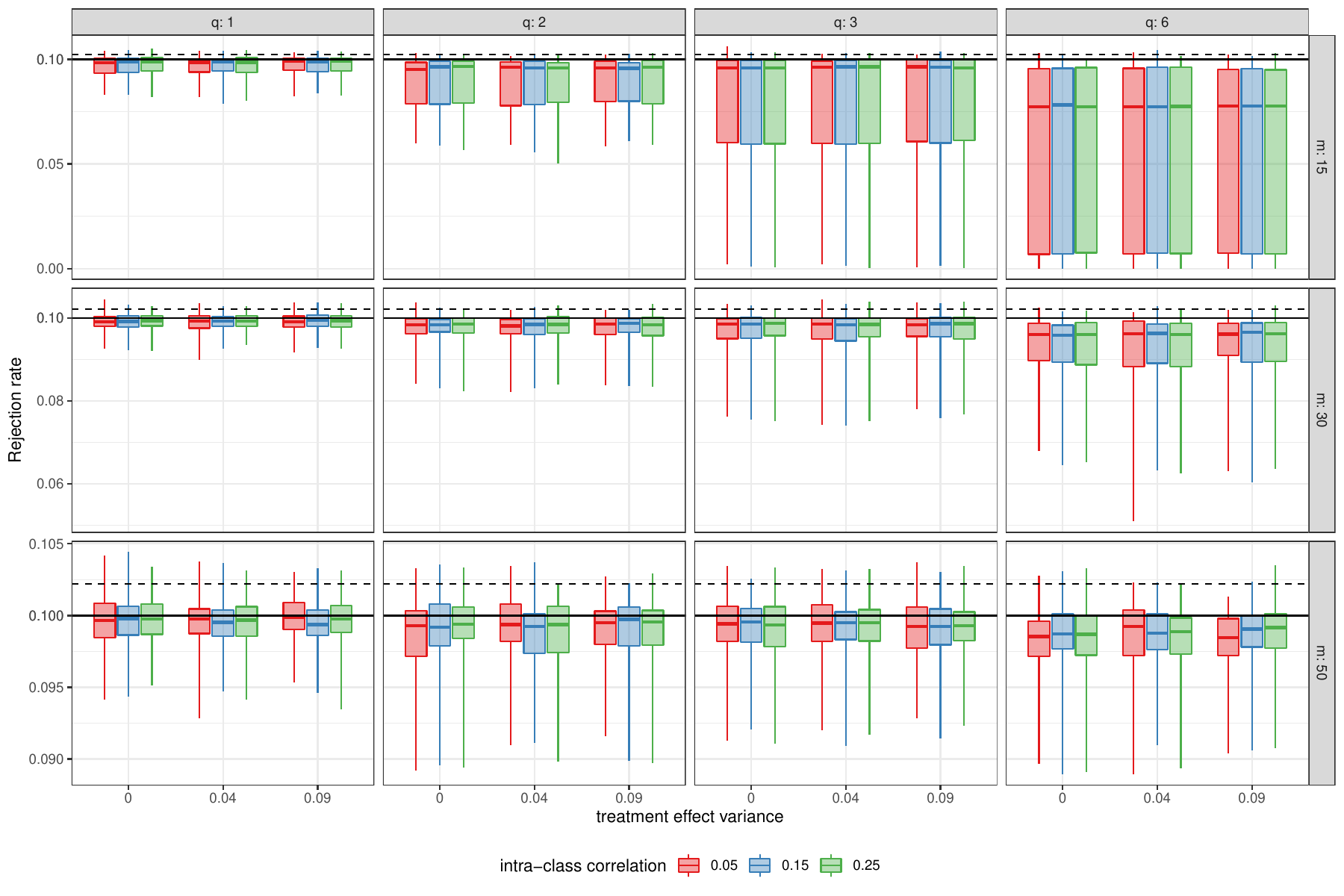} 

}

\caption{Rejection rates of CR2 AHT test, by treatment effect variance and intra-class correlation for $\alpha = .10$.}\label{fig:misspecification_10}
\end{figure}

\end{landscape}

\bibliographystyle{agsm}
\bibliography{bibliography.bib}

\end{document}